\documentclass[prd,aps,twocolumn,preprintnumbers,amsmath,amssymb,nofootinbib,superscriptaddress,notitlepage,floatfix]{revtex4-1}

\usepackage{epsfig}
\usepackage{graphicx}
\usepackage[utf8]{inputenc}
\usepackage{color}
\usepackage{epstopdf}
\usepackage[colorlinks=true,
linkcolor=blue,
breaklinks=true,
urlcolor=blue,
citecolor=blue]{hyperref}

\newcommand{\Rnum}[1]{\uppercase\expandafter{\romannumeral #1\relax}}
\newcommand{\nn}{\nonumber}
\newcommand{\be}{\begin{equation}}
\newcommand{\ee}{\end{equation}}
\newcommand{\bea}{\begin{eqnarray}}
\newcommand{\eea}{\end{eqnarray}}
\newcommand{\ba}{\begin{array}}
\newcommand{\ea}{\end{array}}
\newcommand{\bi}{\begin{itemize}}
\newcommand{\ei}{\end{itemize}}

\newcommand{\mcb}{{\mathcal B}}

\newcommand{\lf}{\left}
\newcommand{\rg}{\right}

\newcommand{\bew}{\begin{widetext}}
\newcommand{\eew}{\end{widetext}}
\newcommand{\landscapeplot}[2]{%
  \resizebox{#1}{!}{\rotatebox{0}{\includegraphics{#2}}}}

\newcommand{\ucas}{\affiliation{University of Chinese Academy of Sciences, Beijing 100049, China}}

\newcommand{\keylab}{\affiliation{State Key Laboratory of Heavy Ion Science and Technology, Institute of Modern Physics, Chinese Academy of Sciences, Lanzhou 730000, China}}

\newcommand{\dongnan}{\affiliation{School of Physics, Southeast University, NanJing 211189, China}}

\newcommand{\peng}{\affiliation{Peng Huanwu Collaborative Center for Research and Education, International Institute for Interdisciplinary and Frontiers,\\ Beihang University, Beijing 100191, China}}
\begin{document}

\title{Strange Two-Photon Exchange in Time-Like Electromagnetic Form Factors}


\author{Xu Cao} \email{caoxu@impcas.ac.cn}
\keylab
\ucas

\author{Yong-Hui Lin} \email{yonghuil@buaa.edu.cn}
\peng

\author{Hai-Qing Zhou} \email{zhouhq@seu.edu.cn}
\dongnan

\date{\today}

\begin{abstract}
  \rule{0ex}{3ex}
    While two-photon exchange (TPE) effects in space-like nucleon electromagnetic form factors have been extensively studied for over two decades, their manifestations in the time-like region have yet to be definitively established. We propose the isospin-violating decay $J/\psi \to \bar{\Sigma}^{0}\Lambda$ and its charge-conjugate process as a sensitive probe of TPE contributions to time-like hyperon electromagnetic form factors.
    This channel provides a clean laboratory for isolating TPE effects through sensitive and robust polarization and spin-correlation observables, which are particularly enhanced by the $\Lambda$-$\Sigma^0$ mass splitting.
    Combining a dispersion-relation analysis of the one-photon-exchange contribution with explicit loop calculations of the TPE amplitudes, we perform a quantitative study of TPE effects in the threshold region. Our results demonstrate that the considered observables can provide unambiguous signatures of TPE contributions and present predictions for their size together with the associated theoretical uncertainties.
\end{abstract}

\maketitle

\section{Introduction}

Nucleons account for the vast majority of visible mass, arising from the intricate non-perturbative dynamics of quarks and gluons \cite{Gross:2022hyw}.
Ever since Hofstadter’s pioneering electron-proton scattering experiments first revealed the proton's finite size in the 1950s \cite{Hofstadter:1955ae,Mcallister:1956ng,Hofstadter:1956qs}, this method has become an essential technique for probing the internal structure of nucleons.
Beyond the Born level, two-photon exchange (TPE) effects in the space-like region \cite{Blunden:2003sp,Guichon:2003qm} have been identified as a leading candidate to resolve the discrepancy between Rosenbluth and polarization transfer measurements of nucleon form factors \cite{Afanasev:2017gsk,Puckett:2017flj}.
The nucleon and $\Delta$ intermediate states dominate the evaluation of TPE effects~\cite{Kondratyuk:2005kk,Blunden:2017nby}, where a partial cancellation is observed between the contributions of the spin-$1/2$ and spin-$3/2$ resonances\cite{Kondratyuk:2007hc,Kondratyuk:2005kk}.
TPE corrections are also relevant in a wide range of precision observables, including parity-violating elastic electron-proton scattering through their interplay with $\gamma Z$-exchange contributions \cite{Zhou:2007hr}, electron-nucleus scattering \cite{Tereshchuk:2026iez}, and inclusive and semi-inclusive deep-inelastic scattering \cite{Klest:2025yik,Lee:2025dsz}. Moreover, TPE effects constitute the dominant theoretical uncertainty in the Lamb shift and hyperfine splitting of muonic hydrogen, stimulating substantial theoretical efforts, including lattice QCD calculations, to reduce this uncertainty \cite{Fu:2022fgh}.

The proton Dirac form factor has been calculated up to next-to-leading order (NLO) in perturbative QCD~\cite{Huang:2024ugd,Chen:2024fhj}, and higher-twist contributions have also been addressed~\cite{Yu:2025jcs}.
Such theoretical evaluations of nucleon and hyperon electromagnetic form factors (EMFFs) have advanced considerably through frameworks like Lattice QCD \cite{Lin:2008mr,Shanahan:2014cga,CSSM:2014knt,Jang:2018djx,Lin:2020rxa,Constantinou:2020hdm}, QCD sum rule \cite{Liu:2009mb}, covariant quark-diquark model \cite{Ramalho:2019koj,Ramalho:2012ad}, chiral perturbation theory \cite{Kubis:2000aa}, Dyson-Schwinger equations \cite{Sanchis-Alepuz:2017mir} and Faddeev equation \cite{Sanchis-Alepuz:2015fcg,Sanchis-Alepuz:2017mir,Liu:2023reo,Cheng:2025yij}.
Unlike nucleons, hyperon EMFFs are experimentally accessible only in the timelike region.

Two-photon exchange (TPE) effects, as well as more general next-to-leading-order (NLO) QED corrections, can be systematically calculated for point-like lepton-pair production in the time-like region. However, for processes involving baryon-antibaryon final states, a complete and model-independent framework for baryonic QED corrections is still lacking, with only a limited number of exploratory studies available to date \cite{Rekalo:2003xa,Rekalo:2003km,Rekalo:2004wa,Gakh:2005hh,Gakh:2005wa,Tomasi-Gustafsson:2005svz,Adamuscin:2007xn,Chen:2008hka}.
This difficulty is further compounded by the intrinsic complexity of the one-photon-exchange (OPE) amplitude in the time-like region, where it becomes complex-valued and suffers from considerable theoretical uncertainties. The absorptive contributions are largely driven by vector-meson resonances \cite{Qian:2022whn,Dai:2023vsw,Cao:2021asd,Xiao:2019qhl,Yan:2023nlb} and final-state interactions \cite{Haidenbauer:2020wyp,Dai:2024lau,Jia:2024ybo,Yang:2024iuc}, which are essential for understanding the oscillatory structures observed in the time-like proton form factors \cite{Bianconi:2015owa}. In addition, the introduction of baryon form factors into loop diagrams leads to significant model dependence and large uncertainties in the predicted interference contributions associated with TPE effects \cite{Chen:2008hka,Zhou:2009xb,Chen:2009ze,Zhou:2010zzt,Zhou:2011yz}.

Yet, despite the rigorous link between these two kinematic regimes provided by dispersion relations \cite{Borisyuk:2008es,Blunden:2017nby,Lin:2021xrc,Lin:2022baj,Lin:2022dyu},
TPE in timelike proton form factors remains poorly understood due to significant theoretical and experimental challenges. Following early attempts based on low-statistics data \cite{Tomasi-Gustafsson:2007vpn,Tomasi-Gustafsson:2004fss}, recent efforts have sought to identify TPE signatures near the proton-antiproton threshold using high-statistics data from BESIII \cite{Xia:2025rio}. A recent phenomenological analysis re-examining the high-statistics BESIII data for at $\psi(3686)$ region found no significant evidence of TPE corrections~\cite{Gao:2026axl}. Crucially, neither of these studies of angular distributions incorporates a complete set of one-loop QED contributions. The lack of generator-level baryonic QED corrections poses significant challenges for data analysis.
Exploiting the self-analyzing nature of hyperon weak decays to measure polarization and correlation observables provides a powerful means of disentangling TPE effects in timelike EMFFs. However, the attempts to search for TPE signatures using these observables at BESIII have yielded no significant signals \cite{BESIII:2019nep}.

In the present study, we propose the $\bar{\Sigma}^{0}\Lambda$ channel in charmonium decay, along with its charge conjugate ($c.c.$) mode, as a sensitive probe for detecting TPE effects.
As an isospin-violating process, the decay proceeds through a purely electromagnetic mechanism. Under this hypothesis, the relative phase between the resonance-mediated process ($e^+ e^- \to \psi \to \Sigma^0 \bar{\Lambda} + c.c.$) and the continuum process ($e^+ e^- \to \gamma^* \to \Sigma^0 \bar{\Lambda} + c.c.$) is expected to be zero~\cite{Korner:1976hv,Claudson:1981fj}. Indeed, within current substantial uncertainties, experimental observations are consistent with this purely electromagnetic hypothesis at $J/\psi$ energy \cite{BESIII:2012xdg,BESIII:2023pfv,BESIII:2023cvk}:
\be
\frac{\sigma_{e^+ e^- \to \gamma^* \to \Sigma^0 \bar{\Lambda} + c.c}/\Gamma_{\psi \to \Sigma^0 \bar{\Lambda} + c.c.}}{\sigma_{e^+ e^- \to \gamma^* \to \mu^+\mu^-}/\Gamma_{\psi \to \mu^+\mu^-}} = 1.02 \pm 0.16
\ee
Consequently, the TPE contribution becomes relatively significant due to the isospin suppression of the charmonium amplitude, with its measurement further facilitated by the high statistics at the charmonium resonances.
Moreover, as will be demonstrated later, the $\Sigma^{0}$ and $\Lambda$ intermediate states do not exhibit the cancellation seen between the nucleon and $\Delta$ in the proton case~\cite{Zhou:2009xb,Chen:2009ze}. Instead, these TPE contributions are enhanced by constructive interference.

Owing to the neutral nature of the $\bar{\Sigma}^{0}\Lambda$ final state, infrared divergences associated with final-state photon emission are absent. This allows the finite TPE contribution to be evaluated separately from the standard real radiative corrections.
This separation circumvents the complexities typically encountered in data interpretation for charged-particle processes.
Consequently, this process serves as an ideal testbed for isolating TPE effects, thereby offering a crucial benchmark for future studies of more complicated baryonic channels.
In fact, even within the OPE approximation, this reaction remains a subject of intense theoretical study, particularly for its role in discriminating between various decay mechanisms~\cite{Zhu:2015bha,BaldiniFerroli:2019abd,Ferroli:2020xnv,Mo:2021asa,Mo:2023wrf,Geng:2023yqo,Rosini:2025wfm,Chen:2025xci,Zhang:2025oks}.


\section{Amplitudes} \label{sec:formula}

\begin{figure}[!hbp]
    \centering
    \includegraphics[width=\linewidth]{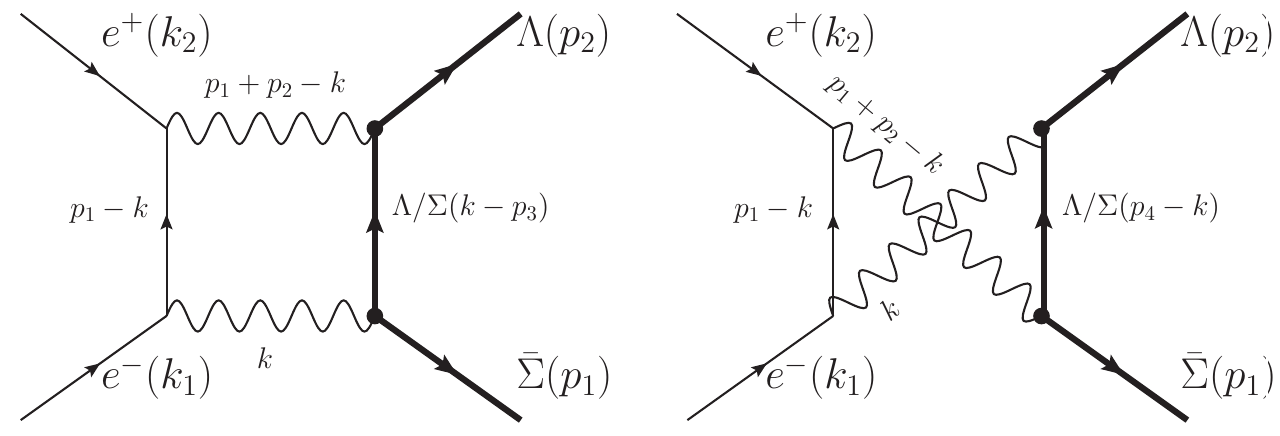}
    \caption{Two-photon-exchange box and crossed-box diagrams for $e^+e^-\to\Lambda\bar\Sigma^0$.  Replacing the external $\Lambda$ and $\bar\Sigma^0$ by $\Sigma^0$ and $\bar\Lambda$, respectively, gives the charge-conjugate channel.  Both $\Lambda$ and $\Sigma^0$ intermediate states are included.}
    \label{fig:TPE}
\end{figure}


We consider the reactions
$e^-(k_1)e^+(k_2)\to Y_1(p_2)\bar Y_2(p_1)$, with
$(Y_1,Y_2)=(\Lambda,\Sigma^0)$ or $(\Sigma^0,\Lambda)$.
The external momenta satisfy $p_1^2=m_1^2$ and $p_2^2=m_2^2$, and we define
\begin{align}
q&=k_1+k_2=p_1+p_2,\notag\\
P&=\frac{p_2-p_1}{2},\qquad
K=\frac{k_1-k_2}{2}.
\end{align}
as well as $2M=m_1+m_2$, $\delta m=m_1-m_2$, and
\be
\tau=\frac{q^2}{4M^2},\quad \xi=\frac{q^2}{\delta m^2},\quad \beta=\sqrt{(1-\tau^{-1})(1-\xi^{-1})}.
\ee
To maintain gauge invariance in the presence of unequal final-state masses, it is convenient to introduce the transverse structures
\be \label{eq:transverse_structures}
\widetilde\gamma_\mu=\gamma_\mu+\frac{\delta m}{q^2}q_\mu,
\qquad
\widetilde P_\mu=P_\mu+\frac{m_1^2-m_2^2}{2q^2}q_\mu.
\ee
They obey $q^\mu\widetilde P_\mu=0$ and, after use of the external Dirac equations, $q^\mu\bar u(p_2)\widetilde\gamma_\mu v(p_1)=0$.
Neglecting the electron mass, the most general vector-current matrix element relevant to the present calculation can then be written as
\bew
\be \label{eq:current}
\mathcal M=-\frac{e^2}{q^2}\bar v(k_2)\gamma_\mu u(k_1)\,\bar u(p_2)
\left[
\widetilde F_1\widetilde\gamma^\mu
+\frac{\widetilde F_2}{2M}i\sigma^{\mu\nu}q_\nu
+F_{2\gamma}\frac{4\gamma\cdot K\,\widetilde P^\mu}{4M^2}
+F_{\delta m}\frac{2\gamma\cdot K\gamma^\mu}{2M}
\right]v(p_1),
\ee
\eew
The generalized invariant amplitudes depend on $q^2$ and, beyond the Born approximation, on the scattering angle $\theta$.
The corresponding generalized Sachs form factors are defined by
\be \label{eq:generalized_sachs}
\widetilde G_M=\widetilde F_1+\widetilde F_2,
\qquad
\widetilde G_E=\widetilde F_1+\tau\widetilde F_2.
\ee
The two structures proportional to $F_{2\gamma}$ and $F_{\delta m}$ have no one-photon counterpart and their relation to the axial parametrization of TPE is given in Appendix~\ref{apdx:formula}.
In particular, the latter is an additional independent structure associated with the non-degenerate $\Lambda$ and $\Sigma^0$ masses.


The Lorentz decomposition in Eq.~\eqref{eq:current} is model independent, whereas the invariant amplitudes receive contributions of different dynamical origins.  We separate the full amplitude as
\be \label{eq:amplitude_decomposition}
\mathcal M=\mathcal M^\gamma+\mathcal M^\psi+\mathcal M^{2\gamma}.
\ee
Here $\mathcal M^\gamma$ denotes OPE continuum contribution, $\mathcal M^\psi$ the one-photon conversion through the charmonium resonance, and $\mathcal M^{2\gamma}$ the TPE contribution.
The former two constitute the Born-level amplitudes $\mathcal M^\gamma+\mathcal M^\psi$, for which $F_{2\gamma}$ and $F_{\delta m}$ vanish.


The real and imaginary parts of form factors in $\mathcal M^\gamma$ are obtained from the dispersion theoretical approach described in Appendix~\ref{apdx:dispersion} to explicitly manifest the resonance--continuum interference.
The part of that construction directly needed in the present amplitude is
\bew
\begin{align}
G_{E/M}^\gamma(t)={}&G_{E/M}(0)+G_{E/M}^{\rm mid}(t)+\frac t{12\pi}\int_{4M_\pi^2}^\infty\frac{d t^\prime}{\pi}\frac{q^3(t^\prime)(F_\pi^{V}(t^\prime))^* T_{E/M}^{\pi\pi}}{t^{\prime 3/2}(t^\prime -t -i\epsilon)}.
\label{eq:continuum_dispersion}
\end{align}
\eew
with $t = q^2$. Here the two-pion contribution controls the low-energy spectral strength and $G_{E/M}^{\rm mid}$ parametrizes the missing intermediate- and high-energy contributions.  Its free parameters are fitted to the measured $e^+e^-\to\Lambda\bar\Sigma^0+\mathrm{c.c.}$ cross sections .
Another essential input is the relative magnitude $R^{\gamma}_{em} = |G_E^{\gamma}/G_M^{\gamma}|$ and phase $\Delta \Phi = {\rm arg} (G_E^{\gamma}/G_M^{\gamma})$ between the electric and magnetic form factors of the OPE amplitude at $J/\psi$. These quantities are identical to those describing the one-photon transition to $J/\psi$ and can be determined from experimental measurements~\cite{BESIII:2023cvk} (see Appendix \ref{apdx:formula}).
The same information is used to constrain the real and imaginary parts of the form factors entering $\mathcal{M}^\psi$, while their overall magnitudes are fixed by the measured branching fraction:
\be
\Gamma_{\psi \to \Sigma^0 \bar{\Lambda} + c.c.} = \frac{M_{\psi}\beta}{12 \pi}\lf( |G_M^{\psi}|^2 + \frac{|G_E^{\psi}|^2}{2\tau} \rg)
\label{eq:M_psi}
\ee
and further normalized to the observed cross section at $J/\psi$, accounting for beam-energy spread and vacuum-polarization effects (see Appendix~\ref{apdx:formula}).



Besides providing corrections to the Born-like structures $\widetilde{F}_1$ and $\widetilde{F}_2$, the TPE amplitude $\mathcal{M}^{2\gamma}$ introduces two new invariant amplitudes, $F_{2\gamma}$ and $F_{\delta m}$.
The diagrams used to calculate TPE amplitudes are shown in Fig.~\ref{fig:TPE}.  The electromagnetic vertex $\gamma^* Y_1 \bar Y_2$ is written as
\be \label{eq:neutral_vertex}
\Gamma_Y^\mu=F_1^Y(k^2) \widetilde \gamma^\mu + \frac{F_2^Y(k^2)}{ 2M}i\sigma^{\mu\nu} k_\nu.
\ee
For the numerical calculation, dipole and Galster-like parametrizations are used for the magnetic and electric form factors, respectively~\cite{Galster:1971kv,Kelly:2004hm,Alberico:2008sz}:
\bea
G_M^Y(k^2)&=&G_M^Y(0)G_D^Y(k^2),\label{eq:dipole}\\
G_E^Y(k^2)&=&-\frac{\mathcal N_Y k^2}{k^2-\Lambda_Y^{\prime 2}}G_D^Y(k^2),\label{eq:Galster}
\eea
where
\be
G_D^Y(k^2)=\frac{1}{(1-k^2/\Lambda_Y^2)^2},
\qquad \mathcal N_Y=0.5.
\ee
The magnetic moments $G^Y_M(0)$ are listed in Table~\ref{tab:F20}.
\begin{table}[!htbp]
    \caption{\label{tab:F20}
      The (transition) magnetic moment $G^Y_M(0) = F^Y_2(0)$ in unit of $\mu_N$ and $\mu_Y$.}
    \begin{ruledtabular}
    \begin{tabular}{l c c}
    $Y$ & $F_2^Y(0)/\mu_N$ & $F_2^Y(0)/\mu_Y$ \\ \hline
    $\Lambda$ & $-0.613\pm0.004$~\cite{ParticleDataGroup:2024cfk} & $-0.729\pm0.005$ \\
    $\Sigma^0\Lambda$ & $1.61\pm0.08$~\cite{ParticleDataGroup:2024cfk,BESIII:2026ubj} & $1.98\pm0.10$ \\
    $\Sigma^0$ & $0.65\pm0.05$~\cite{Geng:2008mf,Flores-Mendieta:2021yzz} & $0.83\pm0.06$
    \end{tabular}
    \end{ruledtabular}
\end{table}

To estimate the model dependence, the cutoff parameters are varied over
\be\label{eq:cutoff}
\Lambda_Y\in[0.84,1.35]~\mathrm{GeV},
\qquad
\Lambda_Y'\in[0.6,1.8]~\mathrm{GeV}.
\ee
These intervals are broadly compatible with Faddeev-equation calculations~\cite{Sanchis-Alepuz:2017mir,Liu:2023reo} and dispersive analyses~\cite{Lin:2022dyu}.
Because $F_1^Y(0)=G_E^Y(0)=0$ for a neutral hyperon and the Pauli coupling carries an explicit photon momentum, the soft region does not generate the usual charge-induced infrared divergence.  The form factors also suppress the ultraviolet region of the hadronic box model.  The opposite signs of $G_M^\Lambda(0)$ and $G_M^{\Sigma^0}(0)$ lead to constructive, rather than cancelling, contributions from the $\Lambda$ and $\Sigma^0$ intermediate states. The loop amplitudes are reduced with \texttt{FeynCalc}~\cite{Mertig:1990an} to four-point Passarino--Veltman functions~\cite{Passarino:1978jh} and evaluated numerically with \texttt{LoopTools}~\cite{Hahn:1998yk}.
All complex-valued form factors are evaluated numerically, and the sub-leading terms quadratic in $\mathcal{M}^{2\gamma}$ are retained in our analysis.

\section{Observables with TPE effects}

The angular distribution in the c.m frame is
\be
\frac{d\sigma}{d\Omega}=\frac{\alpha^2\beta}{4s}D,
\ee
where
\bea \label{eq:angular}
D&=&(1-\xi^{-1})\Bigg\{
|\widetilde G_M|^2(1+\cos^2\theta)
+\frac{1}{\tau}|\widetilde G_E|^2\sin^2\theta\nn\\
&&\hspace{5mm}
+2\tau\beta\operatorname{Re}\!\left[\left(\widetilde G_M-\frac{1}{\tau}\widetilde G_E\right)F_{2\gamma}^*\right]
\sin^2\theta\cos\theta\Bigg\}\nn\\
&&+\frac{2\delta m}{M}\beta\cos\theta\,
\operatorname{Re}\!\left(\widetilde G_M F_{\delta m}^*\right).
\eea
To illustrate the magnitude of TPE effects, the TPE contribution to differential cross sections is isolated by subtracting the OPE component from the full (OPE+TPE) results:
\be
\frac{{d\sigma}^{{\rm OPE+TPE}}-{d\sigma}^{\rm OPE}}{{d\sigma}^{\rm OPE}},
\ee
where the OPE accounts for the continuum contribution and, when applicable, the one-photon conversion via charmonium.

For unpolarized electron and positron beams, the transverse polarization and a non-diagonal spin correlation are
\bew
\bea
P_y&=&\frac{2\sin\theta}{\sqrt\tau D}\Bigg\{
(1-\xi^{-1})\Bigg[\cos\theta\operatorname{Im}(\widetilde G_M\widetilde G_E^*)
+\tau\beta\operatorname{Im}\!\left[\left(\widetilde G_M\cos^2\theta+\widetilde G_E\sin^2\theta\right)F_{2\gamma}^*\right]\Bigg]\nn\\
&&\hspace{12mm}+\tau\beta\operatorname{Im}\!\left[\left(\widetilde G_M-\frac{\delta m}{2M}\frac{\widetilde G_E}{\tau}\right)F_{\delta m}^*\right]\Bigg\},\nn\\
C_{xz}&=&\frac{2\sin\theta}{\sqrt\tau D}\Bigg\{
(1-\xi^{-1})\Bigg[\cos\theta\operatorname{Re}(\widetilde G_M\widetilde G_E^*)
+\tau\beta\operatorname{Re}\!\left[\left(\widetilde G_M\cos^2\theta-\widetilde G_E\sin^2\theta\right)F_{2\gamma}^*\right]\Bigg]\nn\\
&&\hspace{12mm}+\tau\beta\operatorname{Re}\!\left[\left(\widetilde G_M-\frac{\delta m}{2M}\frac{\widetilde G_E}{\tau}\right)F_{\delta m}^*\right]\Bigg\}.
\eea
\eew
The diagonal correlations $C_{xx}$, $C_{yy}$, and $C_{zz}$ are collected in Appendix~\ref{apdx:formula}; all other polarization and correlation components vanish for unpolarized beams.
As $C$-odd observables, the forward-backward asymmetries of $P_y$ and $C_{xz}$ are constructed via the sum $\mathcal{O}(\cos\theta) + \mathcal{O}(-\cos\theta)$, providing a direct measure of TPE effects.

The particularly useful combinations of the two charge-conjugate channels are
\bea \label{eq:null_observables}
 P_y^\Lambda-P_y^{\Sigma}
&=& \frac{4\sin\theta}{D}\sqrt\tau\,\beta\,
\operatorname{Im}\!\left(\widetilde G_MF_{\delta m}^*\right),\\
 C_{xz}^{\Lambda\bar\Sigma}-C_{zx}^{\bar\Lambda\Sigma}
&=& \frac{4\sin\theta}{D}\sqrt\tau\,\beta\,
\operatorname{Re}\!\left(\widetilde G_MF_{\delta m}^*\right),
\eea
which can equivalently be constructed from the two final-state polarizations since charge conjugation implies $P_y^\Sigma=-P_y^{\bar\Sigma}$ and $C_{xz}^{\Lambda\bar\Sigma}=-C_{xz}^{\bar\Lambda\Sigma}$. Both combinations vanish in the one-photon approximation and in the mass-degenerate limit.  Their ratio is therefore a direct phase observable:
\begin{align}
\frac{P_y^\Lambda-P_y^{\Sigma}}{C_{xz}^{\Lambda\bar\Sigma}-C_{zx}^{\bar\Lambda\Sigma}}
=\tan\!\left[\arg \frac{\widetilde G_M}{F_{\delta m}}\right].
\label{eq:phase_ratio}
\end{align}
The ratio does not necessarily inherit the small absolute values of the two vanishing observables, although it becomes numerically unstable when both the numerator and denominator approach zero in the extreme forward and backward directions. A nearly angle-independent ratio indicates that the relative phase of $G_M$ and $F_{\delta m}$ varies only weakly with $\theta$.

\section{numerilcal results}
\label{sec:num}

To briefly summarize the calculation of the form factors discussed in the preceding section,
the OPE form factors entering $\mathcal M^\gamma$ are taken from the dispersion theoretical approach described in Appendix~\ref{apdx:dispersion}.  At the $J/\psi$ energy, the form factors are normalized to the observed resonance contribution, while the relative electric--magnetic phase is fixed by the BESIII polarization measurement summarized in Appendix~\ref{apdx:formula}.  The TPE uncertainty bands are generated by varying the cutoff parameters of the box diagrams over the intervals specified in Eq.~\eqref{eq:cutoff}.


\begin{figure*}[!htbp]
    \centering
    \landscapeplot{0.48\linewidth}{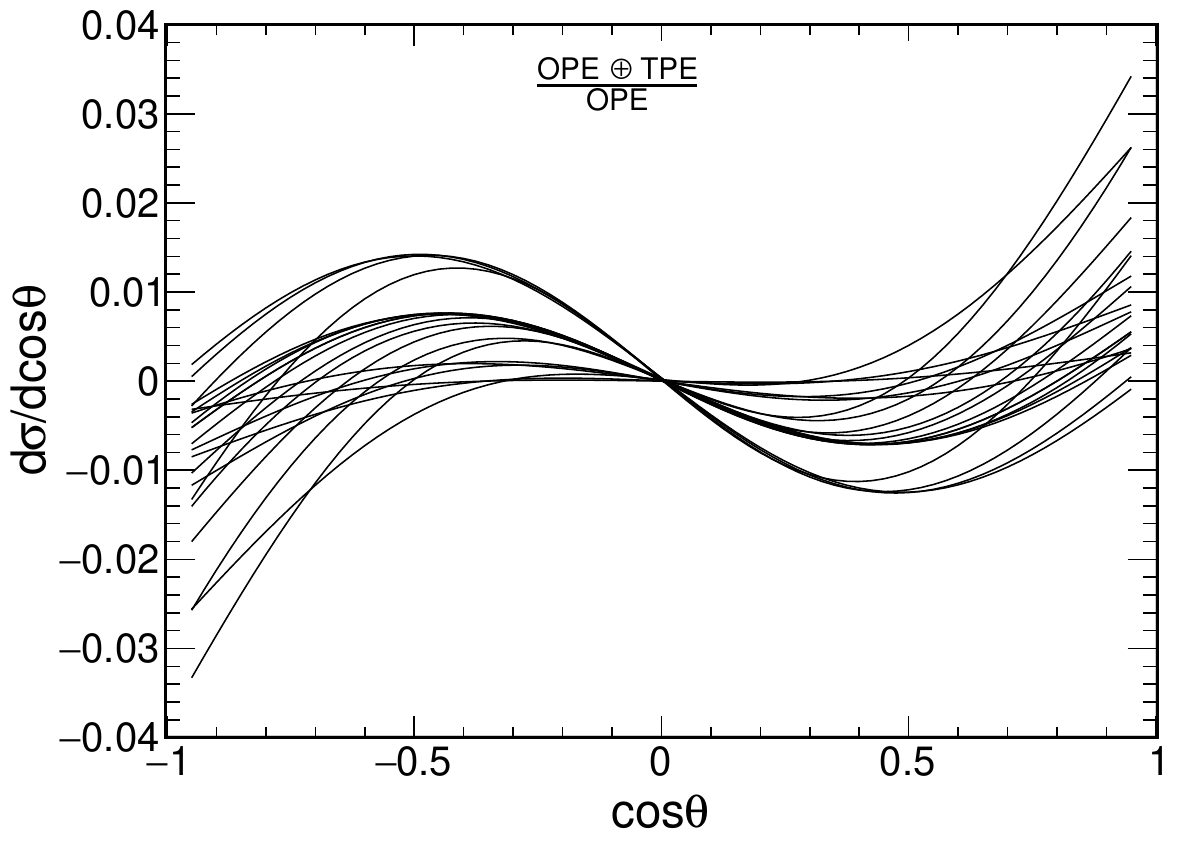}\hfill
    \landscapeplot{0.48\linewidth}{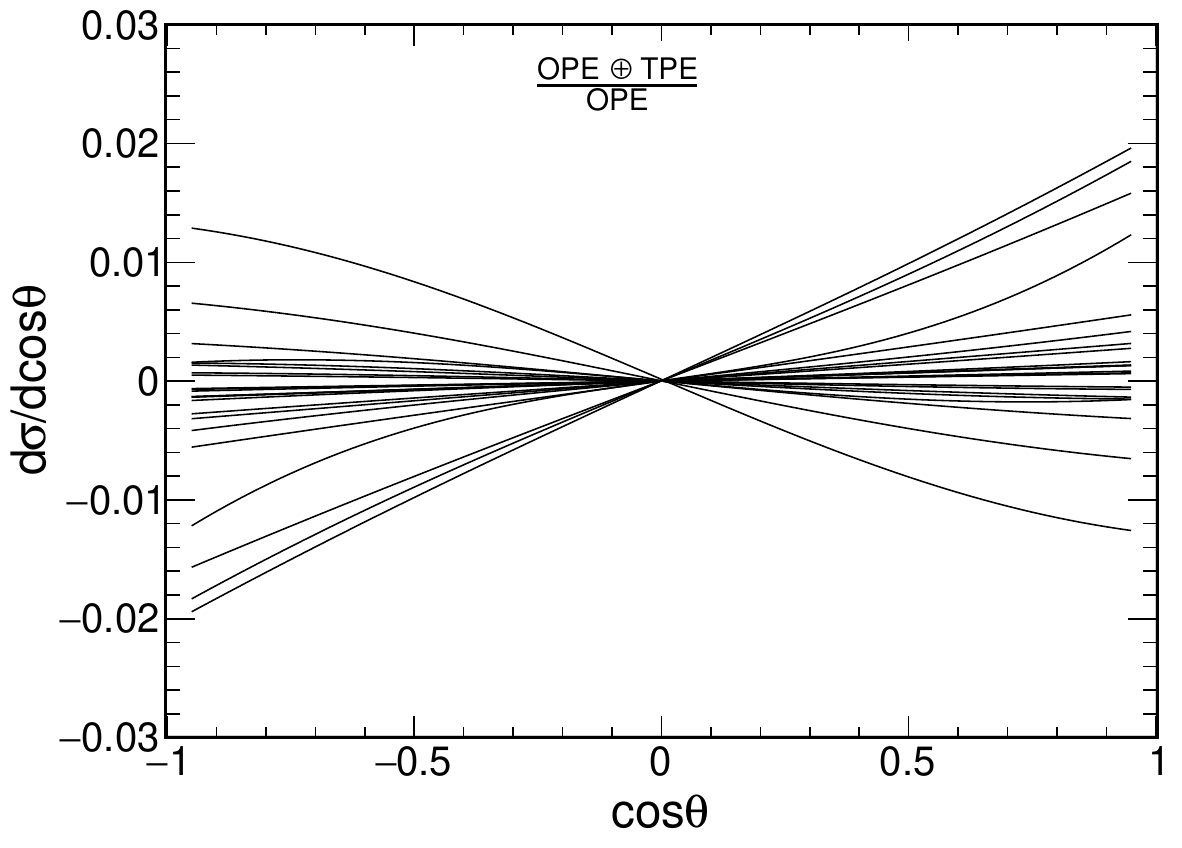}
    \caption{TPE contribution to the differential cross section, defined as the difference between the OPE+TPE and OPE results,  at the $J/\psi$ energy (left) and at $q^2=2.396$~GeV (right).The curves correspond to various cutoff parameters within the range specified in Eq.~\eqref{eq:cutoff}.}
    \label{fig:asydif}
\end{figure*}

Within the central angular regime, the TPE correction to the differential cross section remains below approximately $2\%$ at both the $J/\psi$ resonance and the threshold region, as shown in Fig.~\ref{fig:asydif}.
The small correction to the unpolarized cross section motivates the use of the null combinations in Eq.~\eqref{eq:null_observables}.

\subsection{The $J/\psi$ region}

\begin{figure*}[!htbp]
    \centering
    \landscapeplot{0.48\linewidth}{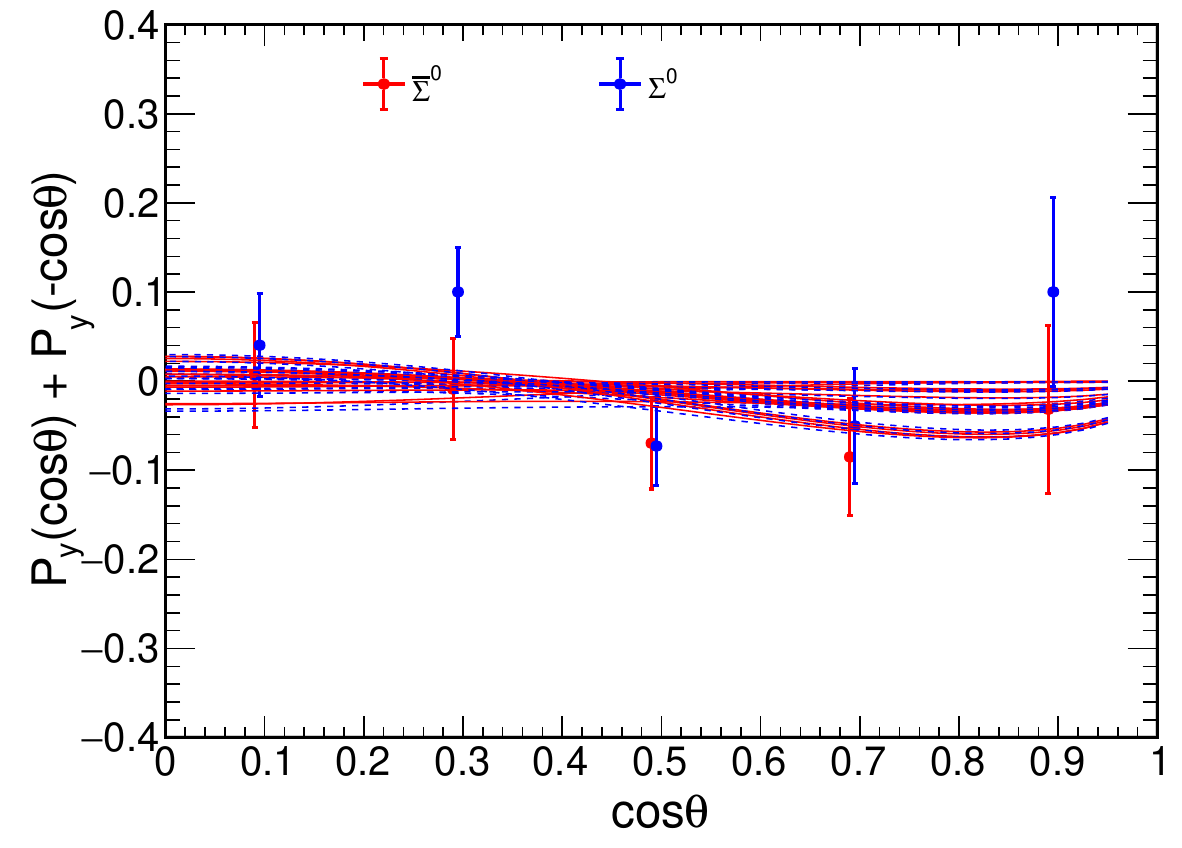}\hfill
    \landscapeplot{0.48\linewidth}{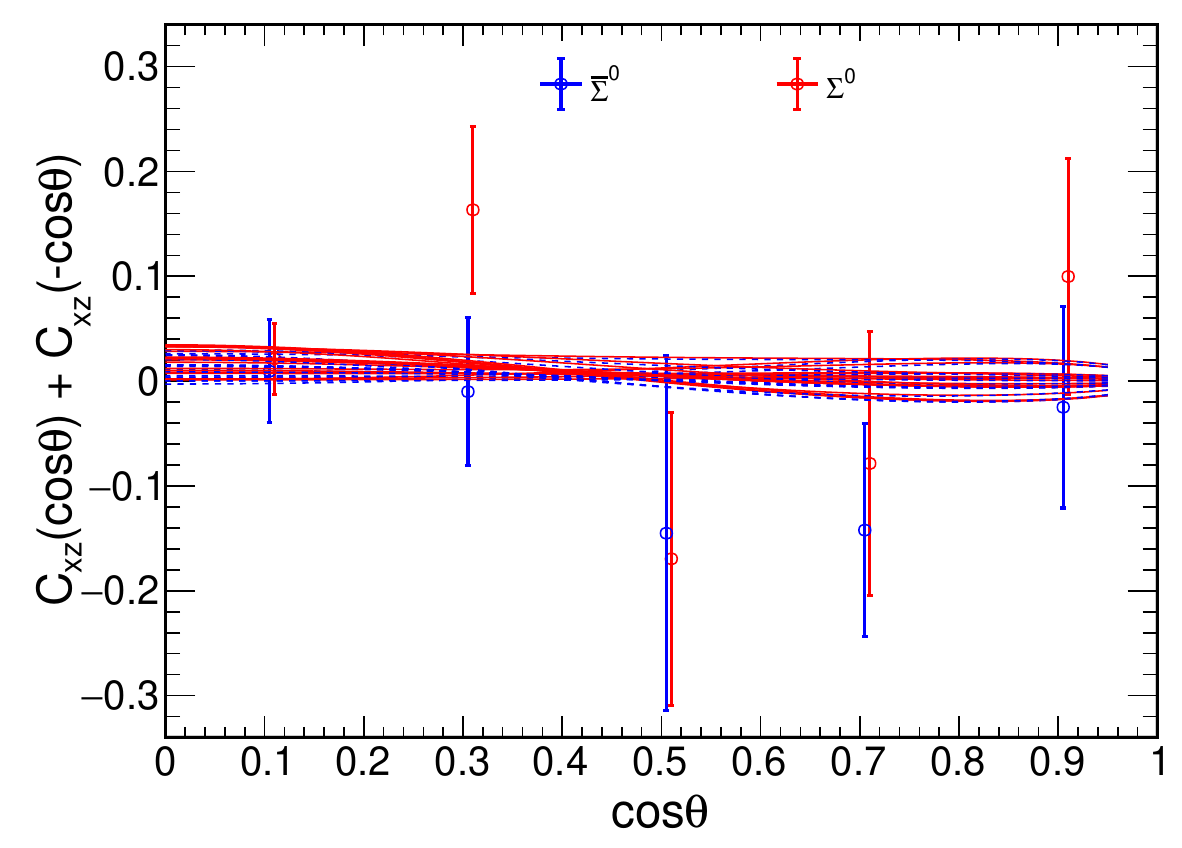}
    \caption{Forward-backward asymmetries for $P_y$ and $C_{xz}$ at $J/\psi$ regime in comparison with BESIII data \cite{BESIII:2023cvk}.}
    \label{fig:corrJpsi}
\end{figure*}


Fig.~\ref{fig:corrJpsi} illustrates the forward-backward asymmetries for $P_y$ and $C_{xz}$, reflecting a marginal but non-vanishing $C$-odd contribution, in agreement with experimental observations.
Specifically, $C_{yy}$ and $C_{zz}$ can help narrow down the allowed parameter space, as certain configurations exhibit sizable forward-backward asymmetries (see Fig.~\ref{fig:diag} in Appendix~\ref{apdx:formula}).

\begin{figure*}[!htbp]
    \centering
    \landscapeplot{0.48\linewidth}{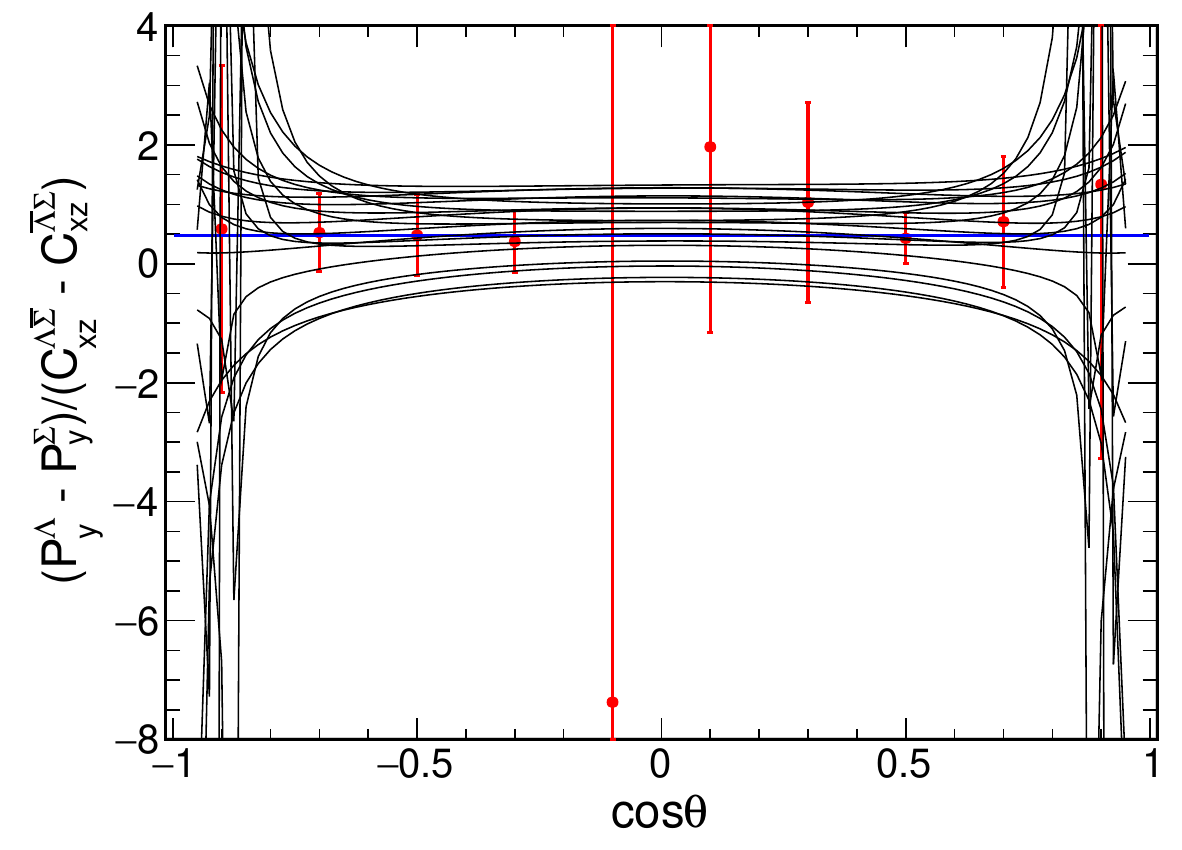}\hfill
    \landscapeplot{0.48\linewidth}{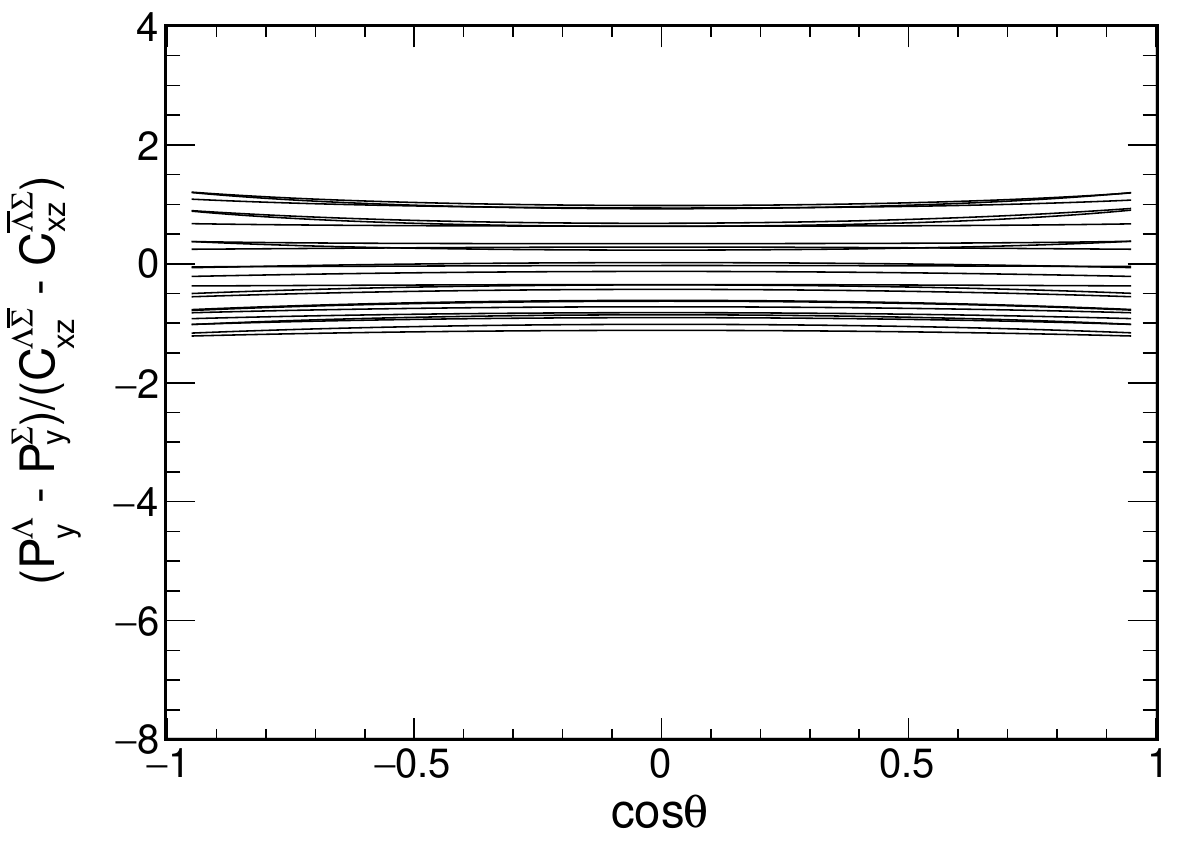}
    \caption{The ratio {$(P_y^\Lambda - P_y^{\Sigma})/(C_{xz}^{\Lambda\bar{\Sigma}} - C_{zx}^{\bar{\Lambda}\Sigma})$} at the $J/\psi$ energy (left), compared with the value deduced from the BESIII analysis~\cite{BESIII:2023cvk}, and at $q^2=2.396$~GeV (right).  The solid blue line in the left panel is a constant fit to the experimental points.}
    \label{fig:asyratio}
\end{figure*}

The ratio $(P_y^\Lambda - P_y^{\Sigma})/(C_{xz}^{\Lambda\bar{\Sigma}} - C_{zx}^{\bar{\Lambda}\Sigma})$ in the left panel of Fig.~\ref{fig:asyratio} remains nearly constant over a broad angular interval.
They isolate the term proportional to $F_{\delta m}$ and are absent in the OPE approximation. Their predicted magnitudes are small but non-vanishing over the central angular region.
Most cutoff choices yield a positive value, while a smaller subset produces a negative value, corresponding to a different quadrant of the relative phase in Eq.~\eqref{eq:phase_ratio}.  The rapid variation near $|\cos\theta|=1$ is not an enhancement of the underlying TPE amplitude: it results from dividing two quantities that vanish simultaneously in the forward and backward limits.
The experimental data are also consistent with an angle-independent behavior, yielding a constant fit of $0.48\pm0.26$.  However, the published extraction did not include TPE in the spin-density matrix~\cite{BESIII:2023cvk}; the comparison should therefore be regarded as indicative rather than as a direct consistency test.

\subsection{The threshold region}

\begin{figure*}[!htbp]
    \centering
    \landscapeplot{0.48\linewidth}{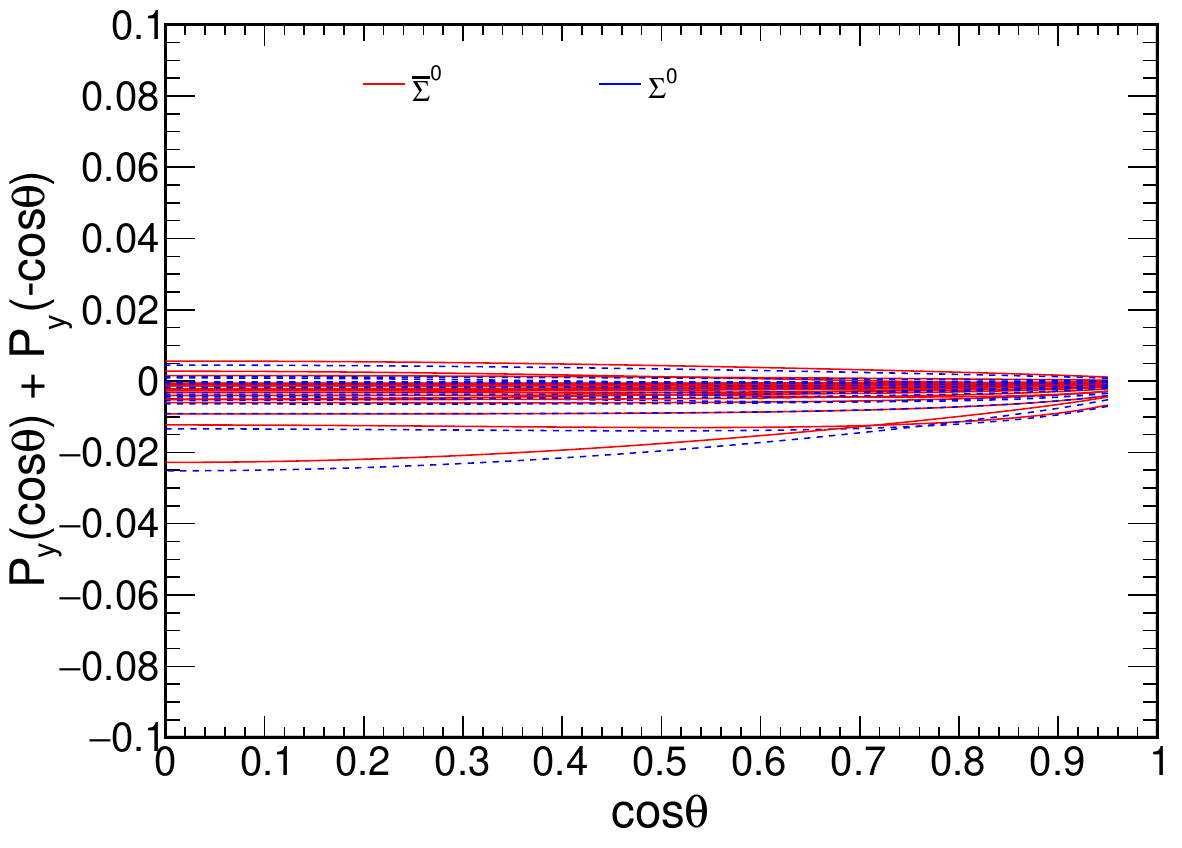}\hfill
    \landscapeplot{0.48\linewidth}{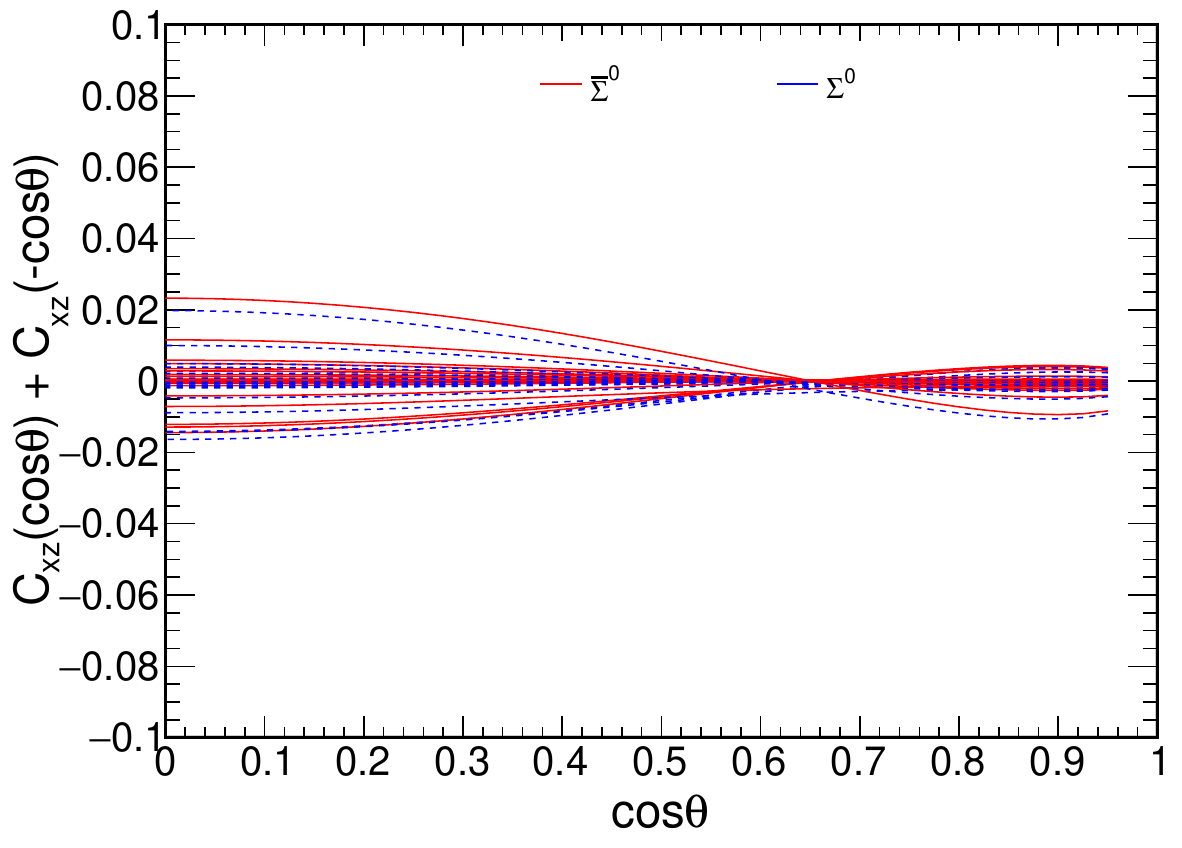}
    \caption{Forward-backward asymmetries for $P_y$ and $C_{xz}$ at $q^2 = 2.396$~GeV.}
    \label{fig:corr24}
\end{figure*}

For $\psi(3686)\to\bar\Sigma^0\Lambda+\mathrm{c.c.}$~\cite{BESIII:2021mus}, the available sample of only several hundred events is insufficient for a complete TPE analysis.  The near-threshold continuum provides an alternative because of its larger production cross section and the prospect of increased luminosity.
At $q^2=2.396$~GeV, $\mathcal M^\psi$ is absent and the dispersive OPE input gives a small electric--magnetic phase difference, $\Delta\Phi_{\bar\Lambda\Sigma^0}=0.026$~rad.  The resulting ratio {$(P_y^\Lambda - P_y^{\Sigma})/(C_{xz}^{\Lambda\bar{\Sigma}} - C_{zx}^{\bar{\Lambda}\Sigma})$}, shown in the right panel of Fig.~\ref{fig:asyratio}, is again nearly constant in the central angular region.  This behavior reflects the weak angular dependence of the phase of $G_M F_{\delta m}^*$ rather than an exact kinematic theorem.

The forward-backward asymmetries for $P_y$ and $C_{xz}$ are shown in Fig.~\ref{fig:corr24}.  With the present cutoff variation, the theoretical band is too broad to establish whether either null observable can be resolved experimentally.  More direct information on the neutral-hyperon form factors, or constraints from diagonal spin correlations, would substantially reduce this uncertainty.

The alternative axial Lorentz basis discussed in Appendix~\ref{apdx:formula} gives the same qualitative behavior.
Such agreement is expected because the two bases are algebraically equivalent, and it serves as a useful cross-check of the numerical implementation of the TPE amplitude projection.

\section{Summary and Perspectives}
\label{sec:sum}

In summary, we have investigated two-photon exchange (TPE) effects in time-like hyperon electromagnetic form factors through the $e^+ e^- \to J/\psi \to \Sigma^0 \bar{\Lambda} + \bar{\Sigma}^0 \Lambda$ process. Our results reveal that the polarization and spin-correlation differences between the hyperon and antihyperon are non-vanishing, arising from the $\Lambda$-$\Sigma^0$ mass splitting. Despite the small magnitude of these symmetry-breaking effects themselves, their ratios can remain sizable, providing a sensitive probe of TPE contributions.

Numerical calculations spanning a wide range of the allowed parameter space--specifically the form-factor cut-offs used to parametrize the high-virtuality behavior of the electromagnetic vertices--indicate that the TPE contribution remains negligible relative to current experimental statistical precision. We emphasize that the charge neutrality of the final state ensures a channel free from infrared divergences, allowing for a separate treatment of TPE and real radiative corrections, which simplifies the interpretation of experimental data. These findings provide a robust benchmark for future experimental and theoretical studies of TPE in the time-like region, particularly in the threshold region.
Our work underscores the unique role of isospin-violating hyperon production as a sensitive laboratory for TPE.

Increasing statistics at BESIII, the upgrade of BEPCII, and  future facilities such as the Super Tau-Charm Facility \cite{Achasov:2023gey} will significantly enhance the precision of hyperon polarization measurements. Such advancements will eventually reach the level required to resolve these subtle higher-order electromagnetic dynamics. Our findings establish a critical baseline for such future explorations and contribute to a deeper understanding of the complex interplay between electromagnetic and hadronic interactions in the timelike region.

\bigskip

\begin{acknowledgments}

We are grateful to Zhe Zhang for useful discussions. This work is supported by the National Key R\&D Program of China under Grant No. 2023YFA1606703, and the National Natural Science Foundation of China  (Grant Nos. 12547111 and 12675097).
The work of YHL was supported in part by the National Science Foundation of China under Grant No. W2543006

\end{acknowledgments}

\bigskip

\appendix
\renewcommand{\thefigure}{\Alph{section}\arabic{figure}}
\setcounter{figure}{0}

\section{Elements of Spin Density Matrix} \label{apdx:formula}
The cross section of $\psi$ contribution reads
\be
    \sigma^{J/\psi} = \frac{12 \pi s}{M_\psi^2} \frac{\Gamma_{ee} \Gamma_{Y_1 \bar{Y}_2 + c.c}}{(q^2-M_\psi^2)^2 + M_\psi^2 \Gamma_\psi^2}
\ee
giving $2.58 \pm 0.21 \, {\rm nb}$ at $J/\psi$ peak with $\mcb (J/\psi \to \Lambda \bar{\Sigma}^0 + {\rm c.c.}) = (2.83 \pm 0.23) \times 10^{-5}$ and the total width $\Gamma_{\psi} = 92.6$ keV \cite{ParticleDataGroup:2024cfk}.
However, the observed cross section corrected by the energy spread and vacuum polarization around $J/\psi$ energy shall be used \cite{BESIII:2018wid,BESIII:2025cpv} and
is estimated to be $\sigma^{J/\psi} =\sigma^{3.08} \times N^{J/\psi}/N^{3.08} = 148 \pm$ 34 pb with $\sigma^{3.08} = 4.4 \pm 0.5 \pm$ 0.4 pb \cite{BESIII:2023pfv} and ratio of events $N^{J/\psi} /N^{3.08} = 33.72 \pm 6.06$ \cite{BESIII:2023cvk}.
Then the $|G_M^{J/\psi}| =0.116 \pm 0.013$ and $|G_E^{J/\psi}| = 0.100 \pm 0.012$ are obtained with $R^{J/\psi}_{em} = |G_E^{J/\psi}/G_M^{J/\psi}| = 0.86 \pm 0.029 \pm 0.015$ \cite{BESIII:2023cvk}.
Their phase difference are measured as $\Delta \Phi_{\bar{\Lambda} \Sigma^0} = 1.011 \pm 0.094 \pm 0.010$ rad, $\Delta \Phi_{\Lambda \bar{\Sigma}^0} = 2.128 \pm 0.094 \pm 0.010$ rad \cite{BESIII:2023cvk}.
The total cross section at $\sqrt{s} = 3.08$~GeV is taken as a measurement of the OPE contribution at the $J/\psi$ region \cite{BESIII:2023pfv}. The interference between the OPE and $J/\psi$ resonant processes can be more rigorously accounted for in the data analysis, given the center-of-mass (c.m.) energy spread of approximately $0.9$~MeV at the $J/\psi$ peak \cite{BESIII:2018wid,BESIII:2025cpv}.

\begin{figure*}[!htbp]
    \centering
    \landscapeplot{0.32\linewidth}{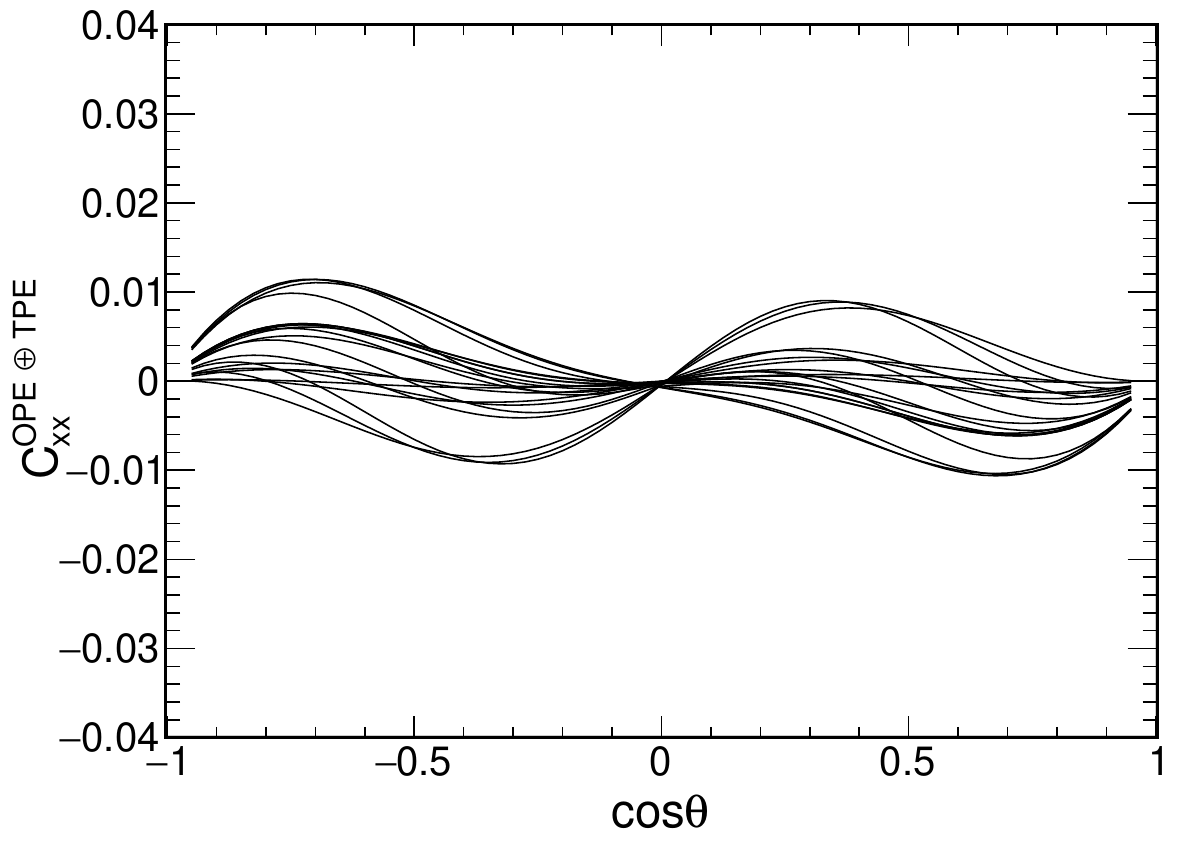}\hfill
    \landscapeplot{0.32\linewidth}{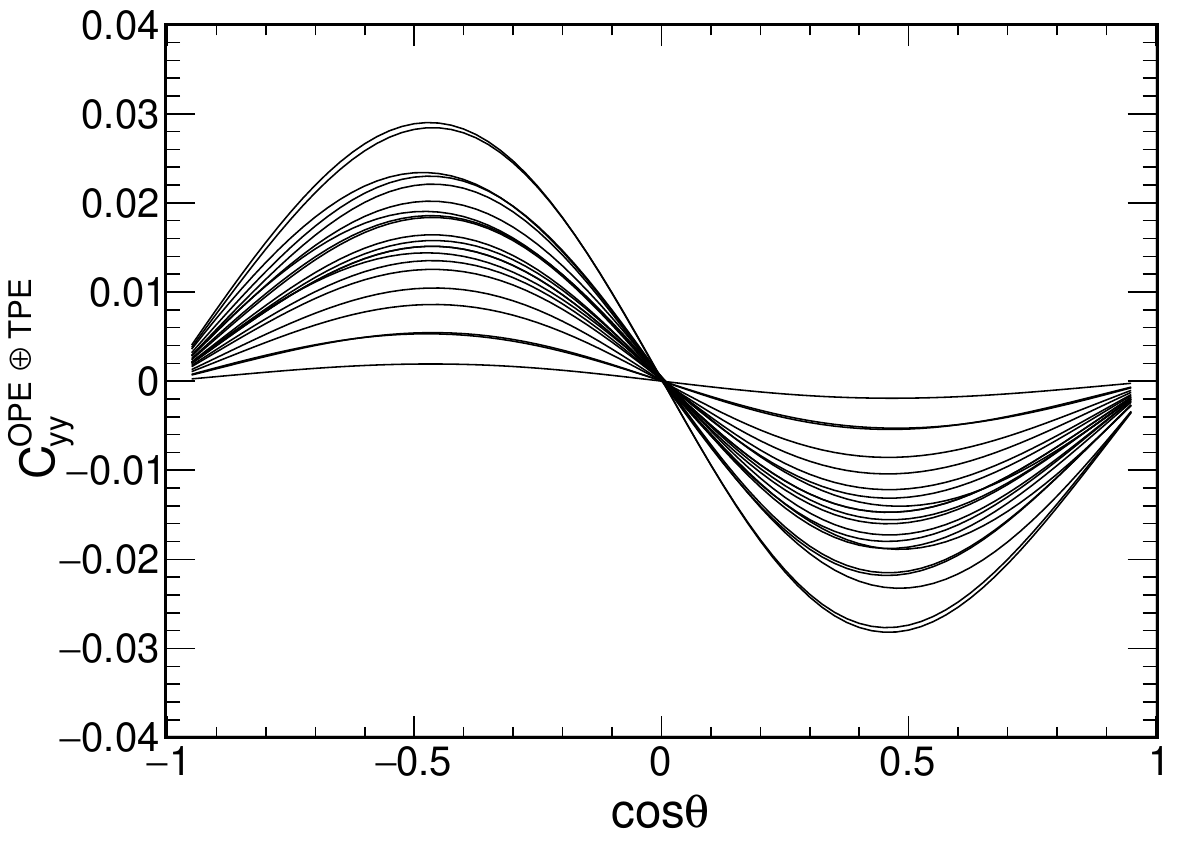}\hfill
    \landscapeplot{0.32\linewidth}{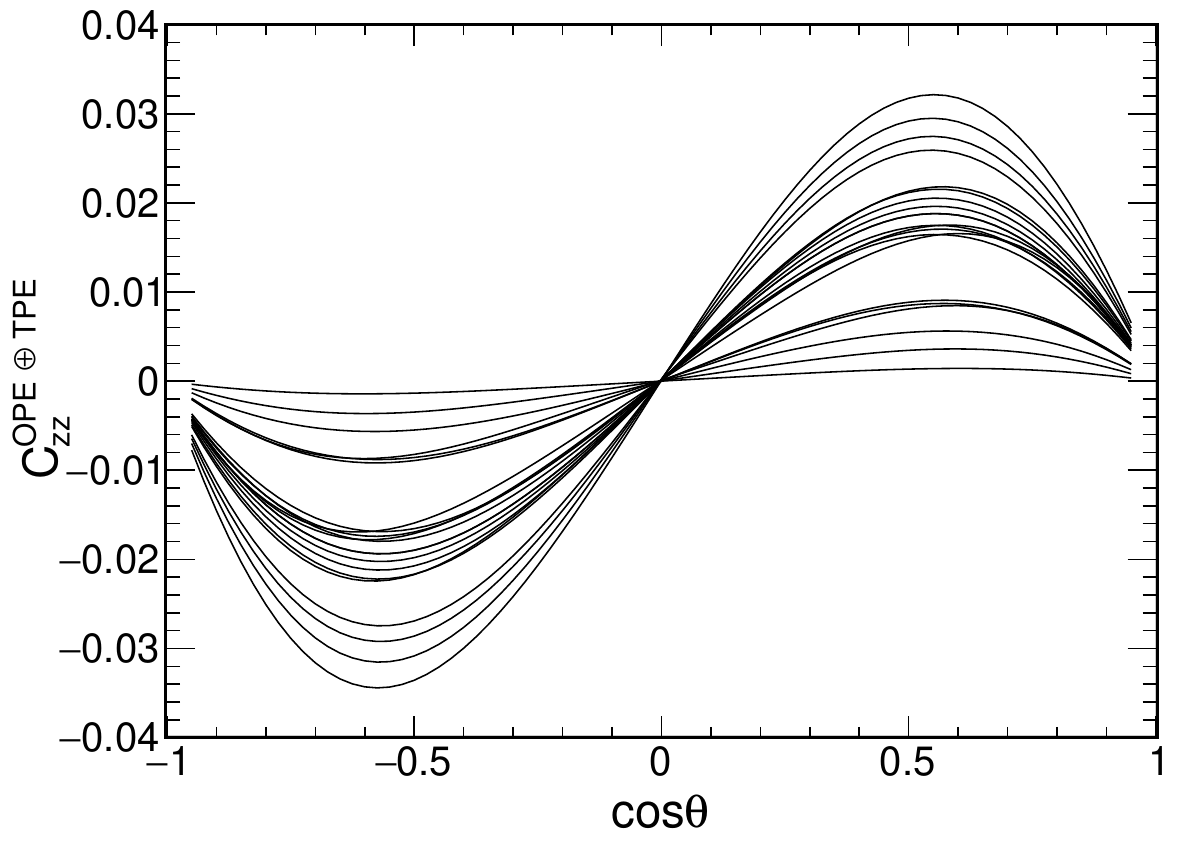}\\[2mm]
    \landscapeplot{0.32\linewidth}{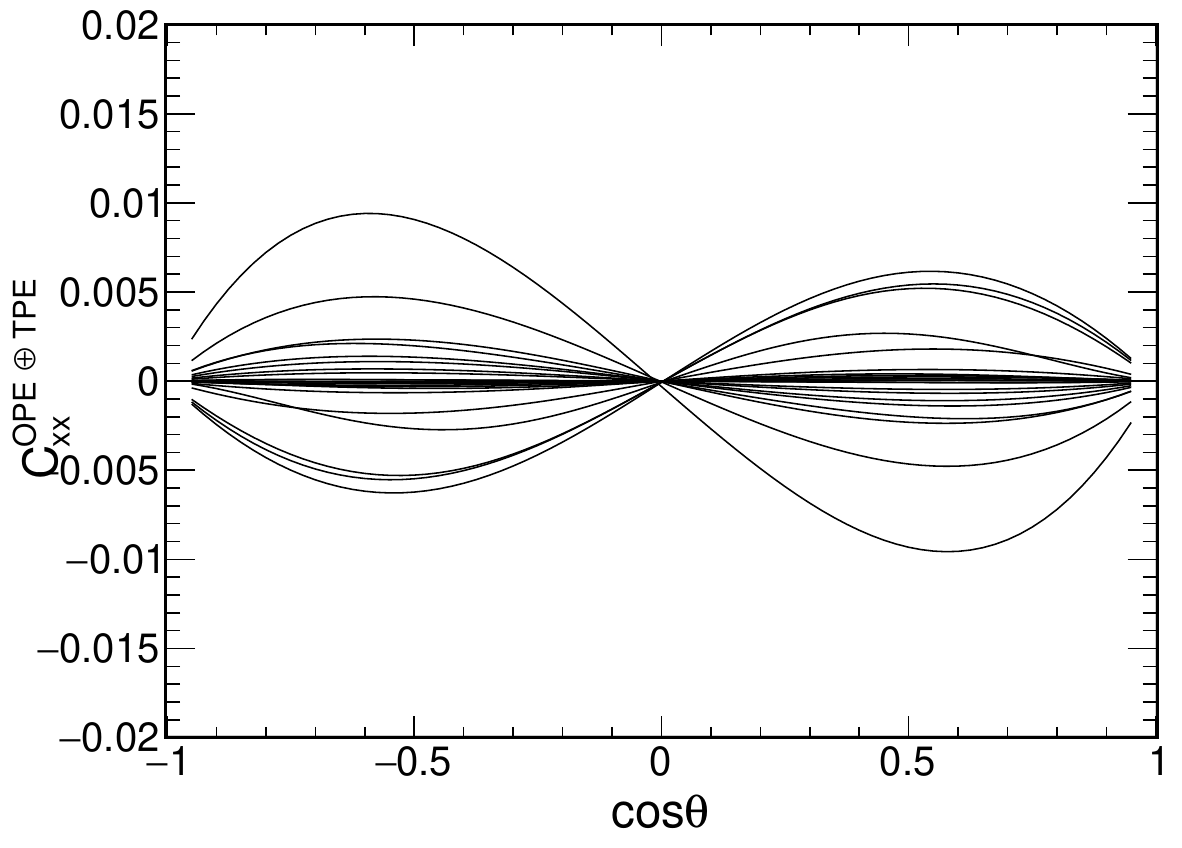}\hfill
    \landscapeplot{0.32\linewidth}{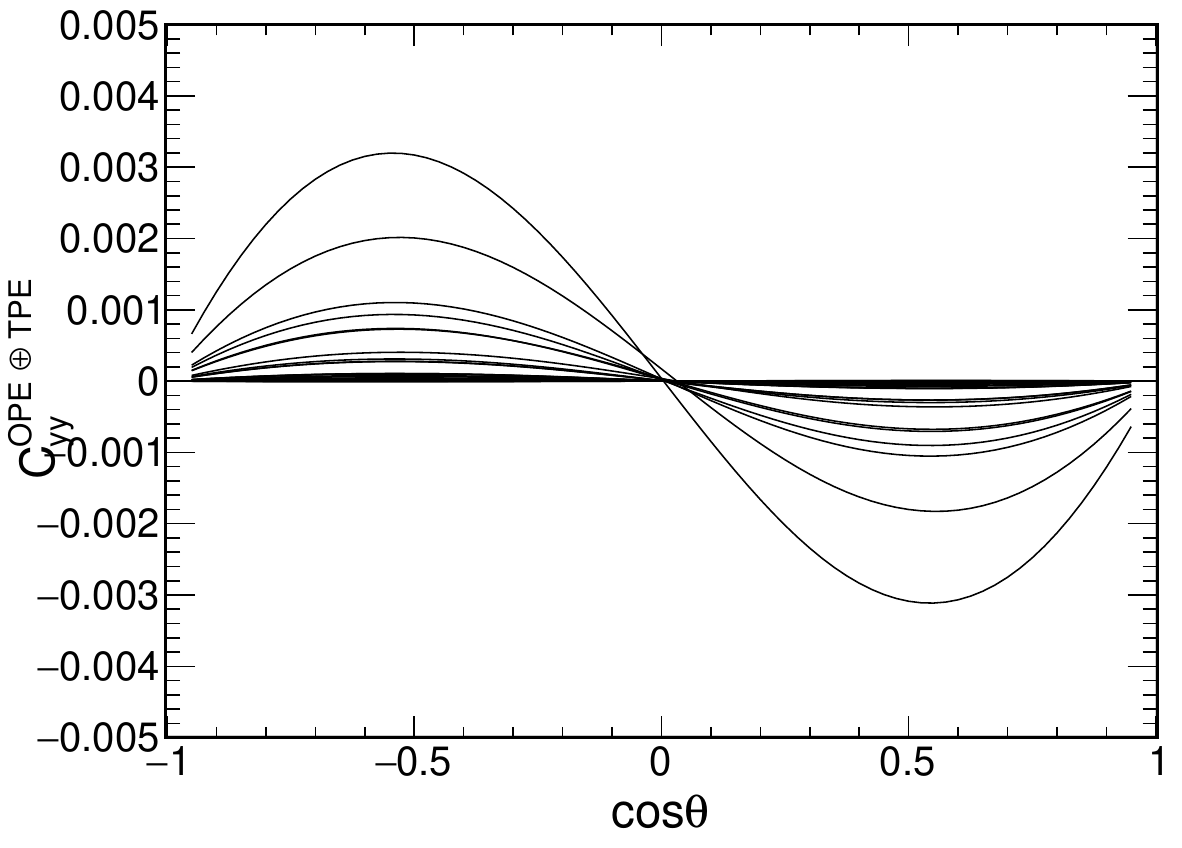}\hfill
    \landscapeplot{0.32\linewidth}{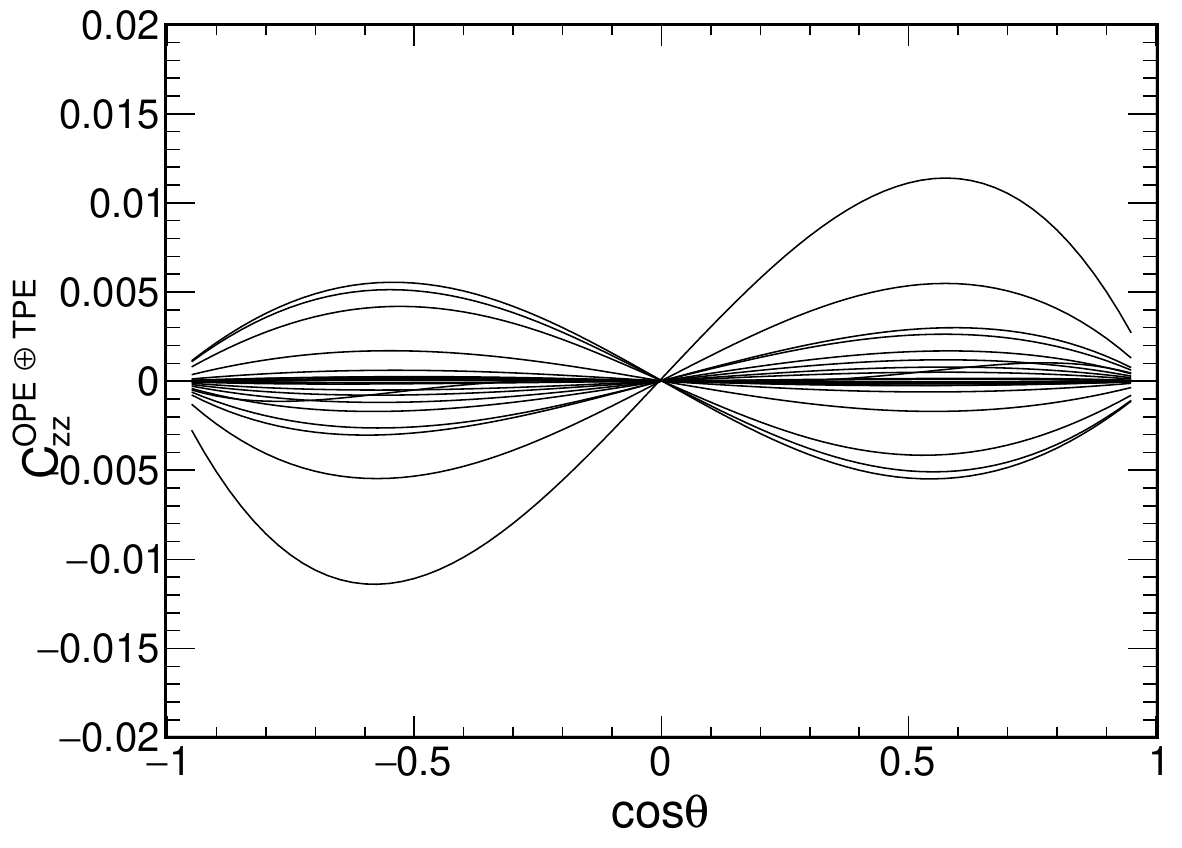}
    \caption{Predicted TPE contributions to the diagonal spin correlations of the hyperon-antihyperon pair at the $J/\psi$ region (upper panels) and at $q^2=2.396$~GeV (lower panels).}
    \label{fig:diag}
\end{figure*}

Spin density matrix of hyperon-anti-hyperon reads as:
\be
D \cdot \left(
\begin{array}{cccc}
 1 & 0 & P_y^{Y_1} & 0 \\
 0 & C_{xx} & 0 &  C_{xz} \\
 P_y^{\bar{Y}_2}  & 0 & C_{yy}  &   0 \\
 0 & C_{zx}  &  0 & C_{zz}  \\
\end{array}
\right),\label{eqn:cmatrixlong}
\ee
$C$-invariance between $e^+ e^- \to  \Lambda \bar{\Sigma}^0$ and $e^+ e^- \to \bar{\Lambda} \Sigma^0$ implies exactly $P_y^\Lambda = P_y^{\bar{\Lambda}}$, $P_y^\Sigma = P_y^{\bar{\Sigma}}$,
$C_{xz}^{\Lambda\bar{\Sigma}} = - C_{xz}^{\bar{\Lambda}\Sigma}$, and
$C_{zx}^{\Lambda\bar{\Sigma}} = - C_{zx}^{\bar{\Lambda}\Sigma}$.
The diagonal correlations of the hyperon-antihyperon pair, satisfying $C_{xx} + C_{yy} + C_{zz} = 1$,  are respectively given by
\bew
\bea
C_{xx}&=&\frac{\sin^2\theta }{D} (1-\xi^{-1}) \lf [|\tilde G_M|^2+
\frac{1}{\tau }|\tilde G_E|^2+ 2{\tau \beta}\cos\theta \mathfrak{Re}(G_M+\frac{1}{\tau }G_E)F_{2\gamma}^*\rg ],
\nonumber \\
C_{yy}&=&\frac{ \sin^2\theta }{D} (1-\xi^{-1}) \lf[ \frac{1}{\tau }|\tilde G_E|^2-
|\tilde G_M|^2 - 2{\tau \beta}\cos\theta \mathfrak{Re}(G_M-\frac{1}{\tau }G_E)F_{2\gamma}^* \rg], \nonumber
\\ C_{zz}&=&\frac{1}{D}\lf \{ (1-\xi^{-1}) \lf[(1+\cos^2\theta )|G_M|^2-
\sin^2\theta \frac{1}{\tau } |G_E|^2 -
2{\tau \beta}\cos\theta \sin^2\theta
\mathfrak{Re}(G_M+\frac{1}{\tau }G_E)F_{2\gamma}^*\rg ] \rg. \nn \\  && \lf. + \frac{2\delta m}{M} \beta \cos\theta \mathfrak{Re}( \tilde G_M {F}^*_{\delta m} ) \rg \} , \nonumber
\label{eq:corrl}
\eea
Predicted TPE contributions to the diagonal spin correlations of the hyperon-antihyperon pair at $J/\psi$ and $q^2=2.396$~GeV are given in Fig. \ref{fig:diag}.
The TPE contribution to $C_{yy}$ is found to be strictly negligible near the threshold.

Alternative choice of Lorentz structure of the $e^+ e^- \to  \Lambda \bar{\Sigma}^0$ reaction reads \cite{Gakh:2005hh,Adamuscin:2007xn}
\bea
\mathcal{M} &=&-\frac{e^2}{q^2}\lf \{ \bar{v}(k_2)\gamma_{\mu}u(k_1)\bar{u}(p_2)\left[ \tilde F_1\gamma_{\mu}+i\frac{\tilde F_2}{2m} \sigma_{\mu\nu} q_{\nu} +{\tilde F}_{\delta m} \frac{2\gamma \cdot K \gamma_{\mu}}{m_1+m_2} \right] v(p_1) + A_{2\gamma} \bar{v}(k_2)\gamma_{\mu} \gamma_5 u(k_1)\bar{u}(p_2) \gamma_{\mu} \gamma_5 v(p_1) \rg\}, \nn
\eea
where the axial parametrization of the TPE matrix element is related to that in Eq.~\eqref{eq:current} by the identity:
\bea
\bar{v}(k_2)\gamma \cdot P u(k_1)\bar{u}(p_2) \gamma \cdot K v(p_1) &=& \frac{t-u}{4} \bar{v}(k_2)\gamma_{\mu} u(k_1)\bar{u}(p_2) \gamma_{\mu} v(p_1) + \frac{s}{4} \bar{v}(k_2)\gamma_{\mu} \gamma_5 u(k_1)\bar{u}(p_2) \gamma_{\mu} \gamma_5 v(p_1) \nn \\ &-&  \frac{\delta m}{2} \bar{v}(k_2)\gamma_{\mu} u(k_1)\bar{u}(p_2) \gamma_{\mu} \gamma \cdot K v(p_1)
\eea
Here, $s$, $u$, $t$ are the Mandelstam variables.
The last term is due to the $\Lambda$-$\Sigma^0$ mass splitting, and vanishes in the mass-degenerate limit \cite{Carlson:2007sp}.
The angular distribution is
\be
D=(1-\xi^{-1})\lf[|\tilde G_M|^2 (1+\cos^2 \theta) + \frac{1}{\tau} |\tilde G_E|^2 \sin^2 \theta \rg] + 4 \beta \cos \theta \mathfrak{Re}\lf[ \tilde G_M (A_{2\gamma}^* + \frac{\delta m}{2M} F^*_{\delta m}) \rg]
\ee
The polarization and correlations of hyperon-antihyperon pair are:
\bea
P_y &=& \frac{2\sin \theta }{D\sqrt{\tau}} \lf \{ (1-\xi^{-1}) \cos\theta \mathfrak{Im} (\tilde G_M \tilde G_E^*) + \beta \mathfrak{Im} (G_E A_{2\gamma}^*) + \tau \beta \mathfrak{Im} ( \tilde G_M-\frac{\delta m}{2M} \frac{\tilde G_E }{\tau}) {F}^*_{\delta m} \rg\} \,,  \\
C_{xz}&=&\frac{2\sin \theta }{D\sqrt{\tau}}  \lf \{ (1-\xi^{-1})  \cos \theta \mathfrak{Re} (\tilde G_M \tilde G_E^*) + \beta \mathfrak{Re} (G_E A_{2\gamma}^*) + \tau \beta \mathfrak{Re}( \tilde G_M-\frac{\delta m}{2M} \frac{\tilde G_E }{\tau} ) {F}^*_{\delta m}  \rg\} , \\
C_{xx}&=&\frac{\sin^2\theta }{D} (1-\xi^{-1}) \lf[|\tilde G_M|^2+ \frac{1}{\tau} |\tilde G_E|^2 \rg], \\
C_{yy}&=&\frac{\sin^2\theta }{ D} (1-\xi^{-1}) \lf[\frac{1}{\tau} |\tilde G_E|^2- |\tilde G_M|^2 \rg], \\
C_{zz}&=&\frac{(1-\xi^{-1})}{D } \lf[ (1+\cos^2\theta )|\tilde G_M|^2-\sin^2\theta \frac{1}{\tau} |\tilde G_E|^2 \rg] + 4 \beta \cos \theta \mathfrak{Re}\lf[ \tilde G_M (A_{2\gamma}^* + \frac{\delta m}{2M} F^*_{\delta m}) \rg],
\eea

\eew

\section{$\Sigma^0$-to-$\Lambda$ transition form factors from the dispersion theoretical approach} \label{apdx:dispersion}

\begin{figure}[htpb]
    \centering
    \includegraphics[width=\linewidth]{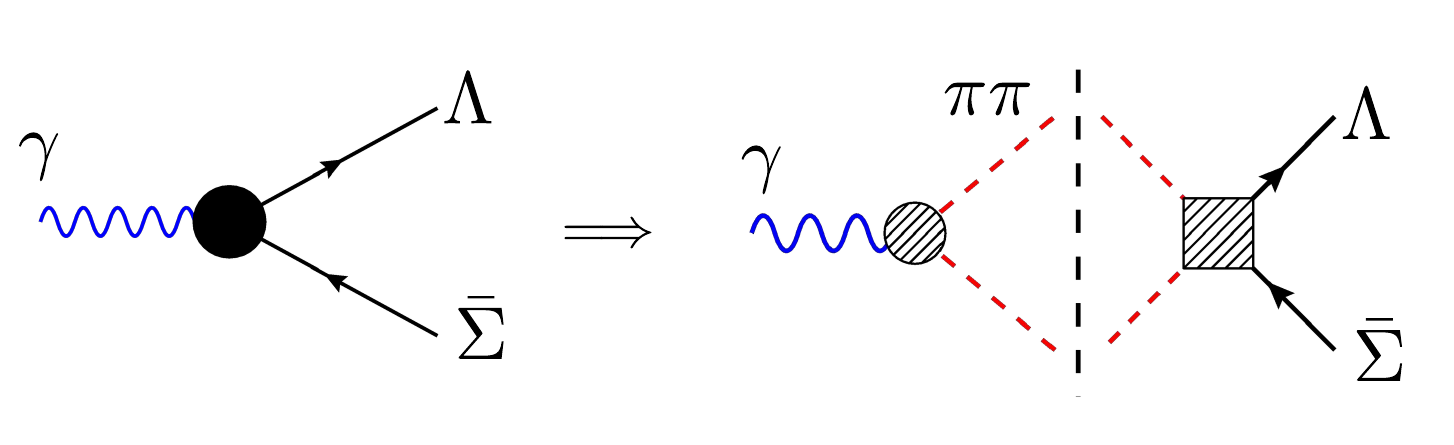}
    \caption{$\Sigma^0$-to-$\Lambda$ transition form factors with the $\pi\pi$ single channel contribution included via dispersion relation formalism.}
    \label{fig: TFF}
\end{figure}

\begin{figure}[htpb]
    \centering
    \includegraphics[width=\linewidth]{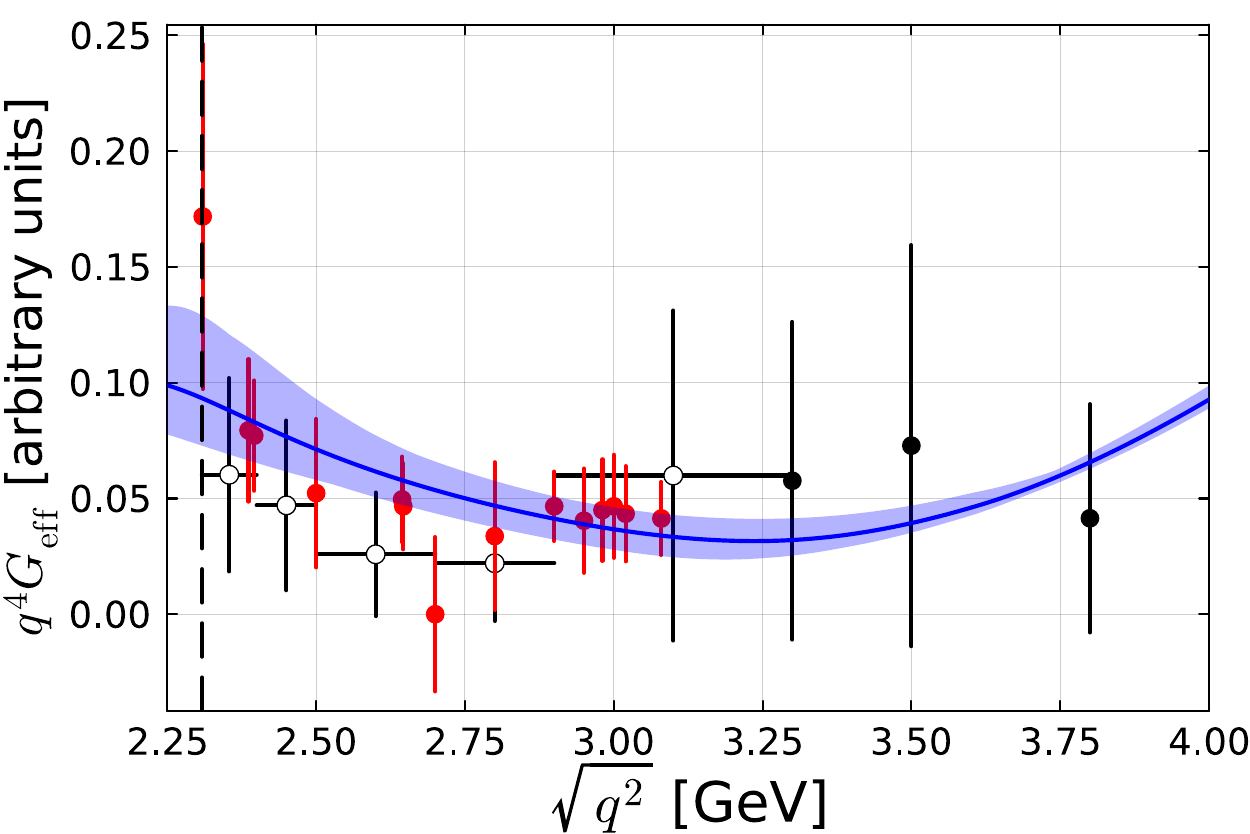}
    \caption{Comparison with the experimental data of the process $e^+e^- \to \Lambda\bar{\Sigma}^0 + c.c$ from Refs.~\cite{BaBar:2007fsu,BESIII:2023pfv}. The black points represent the BaBar data reported in Ref.~\cite{BaBar:2007fsu}, while the BES\Rnum{3} data from Ref.~\cite{BESIII:2023pfv} are shown as red points. Only the filled-disk data points are included in the fit. The vertical dashed line indicates the $\Lambda\bar{\Sigma}^0$ threshold.}
    \label{fig: TFF_fits}
\end{figure}

\begin{table}[b]
    \caption{\label{tab:values}
      {Parameters corresponding to our best fit to the experimental cross-section data for the process $e^+e^- \to \Lambda\bar{\Sigma}^0$ as detailed in the main text. Note that all low-energy constants in three-flavor ChPT are fixed to their central values as used in Ref.~\cite{Lin:2022dyu}, that is, $F_\Phi=100\,{\rm MeV}$, $b_{10}=1.1\,{\rm GeV^{-1}}$, $h_A=2.13$, $D=0.80$ and $F=0.46$. For further details, see Ref.~\cite{Lin:2022dyu}. The $\chi^2/{\rm d.o.f}$ is 0.7. }
    }
    \begin{ruledtabular}
  \begin{tabular}{l c c c c}
$M_V$ (GeV)        &      $g^2$          &       $g_V$          &   $\kappa$  &   $N$          \\ \hline
    2.186$\pm$0.074        &    2.129$\pm$0.326  &    -28.8$\pm$2.7  &    -0.727$\pm$0.052&    1022$\pm$124   \\
     \end{tabular}
  \end{ruledtabular}
  \end{table}

In this section, we present the basic building blocks used to construct the $\Sigma^0$-to-$\Lambda$ transition form factors (TFFs) within the dispersion theoretical framework, which was already investigated in details by Refs.~\cite{Granados:2017cib,Lin:2022dyu}.
From the previous investigation~\cite{Lin:2022dyu}, it is found that the significant contribution to the spectral functions of the electromagnetic transition form factors of $\bar{\Sigma}^0\Lambda$ comes from the two-pion channel.
Then the dispersion relations for the $\Sigma^0$-to-$\Lambda$ TFFs as shown in Fig.~\ref{fig: TFF} can be read as
\begin{align}
    G_{E/M}^\gamma(t)&=G_{E/M}(0)\notag\\
    &+\frac t{12\pi}\int_{4M_\pi^2}^\infty\frac{d t^\prime}{\pi}\frac{q^3(t^\prime)(F_\pi^{V}(t^\prime))^* T_{E/M}^{\pi\pi}}{t^{\prime 3/2}(t^\prime -t -i\epsilon)},
\end{align}
where $F_\pi^V$ is the vector-isovector form factor ($J=I=1$) of the pion and $T_{E/M}^{\pi\pi}$ denotes the $P$-wave $\pi\pi\to \Lambda\bar{\Sigma}^0$ amplitude, which can be constructed with the $SU(3)$ baryon chiral perturbation theory (BChPT) by means of the crossing symmetry.
$q(t)=\sqrt{\lambda(t,M_\pi^2,M_\pi^2)}/2/\sqrt{t}$ is the center-of-mass momentum of the two-pion continuum with $\lambda(a,b,c)=a^2+b^2+c^2-2(ab+bc+ac)$ the K\"all\'en function.
Moreover, we incorporate the two-pion rescattering effect into the next-to-leading-order (NLO) four-point $\pi\pi\to \Lambda\bar{\Sigma}^0$ amplitude using the single-channel Omn\`es formalism. It has been shown that this combination of dispersion relations with BChPT outperforms pure loop calculations when the experimental input is encoded into the Omn\`es representation.
It gives
\begin{align}
    &T_{E/M}(t)=K(t)+\Omega(t)P_{n-1}(t)\notag\\
    &\phantom{xxxx}+\Omega(t)t^n\int_{4M_\pi^2}^\infty \frac{d t^\prime}{\pi}\frac{\sin\delta(t^\prime)K(t^\prime)}{|\Omega(t^\prime)|(t^\prime-t-i\epsilon)t^{\prime n}}.
\end{align}
Here, $\Omega(t)=\exp\left(t\int_{4M_\pi^2}^{\infty}\frac{d t^\prime}{\pi}\frac{\delta(t^\prime)}{t^\prime(t^\prime-t-i\epsilon)}\right)$ is the $P$-wave $I=1$ Omn\`es function for the two-pion channel which is completely determined by the $P$-wave two-pion phase shift $\delta(t)$. $K(t)$ represents the non-polynomial part of the four-point amplitude of $\pi\pi\to \Lambda\bar{\Sigma}^0$, calculated up to NLO in three-flavor ChPT, following Refs.~\cite{Granados:2017cib,Lin:2022dyu}. The term $P_{n-1}$ denotes subtraction polynomial of degree $n-1$, which is identified with the polynomial part of our $\pi\pi\to \Lambda\bar{\Sigma}^0$ amplitude.
In our case, $n=1$. All explicit expressions and numerical details can be found in the aforementioned references.

To apply the above dispersion description for the $\Sigma^0$-to-$\Lambda$ TFFs to the Born amplitude of the process $e^+e^- \to \Lambda\bar{\Sigma}^0$ in the energy region of interest, the TFFs must be calibrated against relevant experimental data. To this end, we adopt the phenomenological strategy developed by Gari and Kr\"umpelmann, originally introduced to study the electromagnetic structure of the nucleon in Ref.~\cite{Gari:1984ia}, in order to model the missing spectral strength in the intermediate energy region. Specifically, we have
\begin{align}
    F_{1, \text{mid}}^{V}(t)&=\left(\frac{M_V^2}{M_V^2-i M_V \Gamma(t)-t}-1\right)g_V F_1(t), \notag\\
    F_{2, \text{mid}}^{V}(t)&=\left(\frac{M_V^2}{M_V^2-i M_V \Gamma(t)-t}-1\right)g_V\kappa F_2(t).
\end{align}
Here, we introduce an energy-dependent width $\Gamma(t)$ to mimic the effective unstable vector meson contribution, which is defined by $\Gamma(t)=g^2/M_V p^3/\sqrt{t}\theta(t-4M_\pi^2)$ with $p=\sqrt{\lambda(t^2,M_\pi^2,M_\pi^2)}/2/\sqrt{t}$ and a dimensionless coupling constant $g$.
$F_1$ and $F_2$ are two phenomenological form factors that contain both the low-energy meson dynamics~\cite{Gari:1984pq} and the high-energy asymptotic quark behavior predicted by perturbative QCD (pQCD)~\cite{Lepage:1980fj}, and are formulated as follows:
\begin{align}
    &F_1(t)=\begin{cases}\frac{1}{1+\gamma_1 t}, & \text{if } t \leq \gamma_1^{-1} \\
                                   N_1\frac{1}{(1+\gamma_1  \hat{t})(1+\gamma_2 \hat{t})},  & \text{if } \gamma_1^{-1} < t \le \gamma_2^{-1}  \\
                                   N_2\left[\frac{1}{t\log\left({t}/{\Lambda_{\rm QCD}^2}\right)}\right]^2,  & \text{if } \gamma_2^{-1} < t
                       \end{cases}\\
    &F_2(t)=\begin{cases}\frac{1}{1+\gamma_1 t}, & \text{if } t \leq \gamma_1^{-1} \\
        N_1^\prime\frac{1}{(1+\gamma_1  \hat{t})(1+\gamma_2 \hat{t})^2},  & \text{if } \gamma_1^{-1} < t \le \gamma_2^{-1}  \\
        N_2^\prime\frac1t\left[\frac{1}{t\log\left({t}/{\Lambda_{\rm QCD}^2}\right)}\right]^2,  & \text{if } \gamma_2^{-1} < t
    \end{cases}
\end{align}
where $\hat{t}=t\log\left((1+\gamma_2 t)/\Lambda_{\rm QCD}^2/\gamma_2\right)/\log\left(1/\Lambda_{\rm QCD}^2/\gamma_2\right)$. For simplicity, we take $\gamma_1^{-1}=0.8^2\ \mathrm{GeV^{2}}$ and $\gamma_2=3.6^2\ \mathrm{GeV^{2}}$, while $\Lambda_{\mathrm{QCD}}=341\ \mathrm{MeV}$ is chosen as the typical three-flavor value.
The normalization factors $N_1^{(\prime)}$ and $N_2^{(\prime)}$ are chosen to ensure that the piecewise-defined functions $F_1$ and $F_2$ are continuous at their connecting points.
Then the corresponding Sachs form factors are given by
\begin{align}
    &G_E^{\text{mid}}=F_{1, \text{mid}}^{V}+\tau F_{2, \text{mid}}^{V},\\
    &G_M^{\text{mid}}=F_{1, \text{mid}}^{V}+ F_{2, \text{mid}}^{V},
\end{align}
with $\tau\equiv t/(m_\Sigma+m_\Lambda)^2$.
And finally, the $\Sigma^0$-to-$\Lambda$ TFFs used to calculate the Born amplitude of $e^+e^- \to \Lambda\bar{\Sigma}^0$ are expressed as
\begin{align}
    &G_{E/M}^\gamma(t)=G_{E/M}(0)\notag\\
    &\phantom{xx}+\frac t{12\pi}\int_{4M_\pi^2}^\infty\frac{d t^\prime}{\pi}\frac{q^3(t^\prime)(F_\pi^{V}(t^\prime))^* T_{E/M}^{\pi\pi}}{t^{\prime 3/2}(t^\prime -t -i\epsilon)}+G_{E/M}^{\text{mid}}(t).
\end{align}
The free parameters $M_V$, $g$, $g_V$, and $\kappa$ are determined by fitting to the experimental cross-section data for the process $e^+e^- \to \Lambda\bar{\Sigma}^0$ reported in Refs.~\cite{BaBar:2007fsu,BESIII:2023pfv} and $\Delta \Phi_{\Lambda \bar{\Sigma}^0}$ in Refs.\cite{BESIII:2023cvk}.

The best solution is presented in Fig.~\ref{fig: TFF_fits} comparing with the experimental data given in Ref.~\cite{BaBar:2007fsu,BESIII:2023pfv}. Note that we analyze the cross section data by translating them to the effective form factors with arbitrary unit, that is defined by
\begin{equation}
    G_{\rm eff}(t)\equiv \sqrt{\frac{2\tau|G_M^\gamma(t)|^2+|G_E^\gamma(t)|^2}{2\tau+1}}=\sqrt{\frac{2\tau}{2
    \tau+1}\frac{2t\sigma^B(t)}{4\pi\alpha^2\beta}}, 
\end{equation}
In addition, two data points for the magnitude and relative phase of the transition form factor at $\sqrt{q^2}=m_{J/\psi}$ are included in our fit.
The resulting parameters are given in Tab.~\ref{tab:values}.

\clearpage
\bibliography{NEFF_all}

\begin{thebibliography}{98}%
\makeatletter
\providecommand \@ifxundefined [1]{%
 \@ifx{#1\undefined}
}%
\providecommand \@ifnum [1]{%
 \ifnum #1\expandafter \@firstoftwo
 \else \expandafter \@secondoftwo
 \fi
}%
\providecommand \@ifx [1]{%
 \ifx #1\expandafter \@firstoftwo
 \else \expandafter \@secondoftwo
 \fi
}%
\providecommand \natexlab [1]{#1}%
\providecommand \enquote  [1]{``#1''}%
\providecommand \bibnamefont  [1]{#1}%
\providecommand \bibfnamefont [1]{#1}%
\providecommand \citenamefont [1]{#1}%
\providecommand \href@noop [0]{\@secondoftwo}%
\providecommand \href [0]{\begingroup \@sanitize@url \@href}%
\providecommand \@href[1]{\@@startlink{#1}\@@href}%
\providecommand \@@href[1]{\endgroup#1\@@endlink}%
\providecommand \@sanitize@url [0]{\catcode `\\12\catcode `\$12\catcode
  `\&12\catcode `\#12\catcode `\^12\catcode `\_12\catcode `\%12\relax}%
\providecommand \@@startlink[1]{}%
\providecommand \@@endlink[0]{}%
\providecommand \url  [0]{\begingroup\@sanitize@url \@url }%
\providecommand \@url [1]{\endgroup\@href {#1}{\urlprefix }}%
\providecommand \urlprefix  [0]{URL }%
\providecommand \Eprint [0]{\href }%
\providecommand \doibase [0]{http://dx.doi.org/}%
\providecommand \selectlanguage [0]{\@gobble}%
\providecommand \bibinfo  [0]{\@secondoftwo}%
\providecommand \bibfield  [0]{\@secondoftwo}%
\providecommand \translation [1]{[#1]}%
\providecommand \BibitemOpen [0]{}%
\providecommand \bibitemStop [0]{}%
\providecommand \bibitemNoStop [0]{.\EOS\space}%
\providecommand \EOS [0]{\spacefactor3000\relax}%
\providecommand \BibitemShut  [1]{\csname bibitem#1\endcsname}%
\let\auto@bib@innerbib\@empty
\bibitem [{\citenamefont {Gross}\ \emph {et~al.}(2023)\citenamefont {Gross}
  \emph {et~al.}}]{Gross:2022hyw}%
  \BibitemOpen
  \bibfield  {author} {\bibinfo {author} {\bibfnamefont {F.}~\bibnamefont
  {Gross}} \emph {et~al.},\ }\href {\doibase 10.1140/epjc/s10052-023-11949-2}
  {\bibfield  {journal} {\bibinfo  {journal} {Eur. Phys. J. C}\ }\textbf
  {\bibinfo {volume} {83}},\ \bibinfo {pages} {1125} (\bibinfo {year}
  {2023})},\ \Eprint {http://arxiv.org/abs/2212.11107} {arXiv:2212.11107
  [hep-ph]} \BibitemShut {NoStop}%
\bibitem [{\citenamefont {Hofstadter}\ and\ \citenamefont
  {McAllister}(1955)}]{Hofstadter:1955ae}%
  \BibitemOpen
  \bibfield  {author} {\bibinfo {author} {\bibfnamefont {R.}~\bibnamefont
  {Hofstadter}}\ and\ \bibinfo {author} {\bibfnamefont {R.~W.}\ \bibnamefont
  {McAllister}},\ }\href {\doibase 10.1103/PhysRev.98.217} {\bibfield
  {journal} {\bibinfo  {journal} {Phys. Rev.}\ }\textbf {\bibinfo {volume}
  {98}},\ \bibinfo {pages} {217} (\bibinfo {year} {1955})}\BibitemShut
  {NoStop}%
\bibitem [{\citenamefont {Mcallister}\ and\ \citenamefont
  {Hofstadter}(1956)}]{Mcallister:1956ng}%
  \BibitemOpen
  \bibfield  {author} {\bibinfo {author} {\bibfnamefont {R.~W.}\ \bibnamefont
  {Mcallister}}\ and\ \bibinfo {author} {\bibfnamefont {R.}~\bibnamefont
  {Hofstadter}},\ }\href {\doibase 10.1103/PhysRev.102.851} {\bibfield
  {journal} {\bibinfo  {journal} {Phys. Rev.}\ }\textbf {\bibinfo {volume}
  {102}},\ \bibinfo {pages} {851} (\bibinfo {year} {1956})}\BibitemShut
  {NoStop}%
\bibitem [{\citenamefont {Hofstadter}(1956)}]{Hofstadter:1956qs}%
  \BibitemOpen
  \bibfield  {author} {\bibinfo {author} {\bibfnamefont {R.}~\bibnamefont
  {Hofstadter}},\ }\href {\doibase 10.1103/RevModPhys.28.214} {\bibfield
  {journal} {\bibinfo  {journal} {Rev. Mod. Phys.}\ }\textbf {\bibinfo {volume}
  {28}},\ \bibinfo {pages} {214} (\bibinfo {year} {1956})}\BibitemShut
  {NoStop}%
\bibitem [{\citenamefont {Blunden}\ \emph {et~al.}(2003)\citenamefont
  {Blunden}, \citenamefont {Melnitchouk},\ and\ \citenamefont
  {Tjon}}]{Blunden:2003sp}%
  \BibitemOpen
  \bibfield  {author} {\bibinfo {author} {\bibfnamefont {P.~G.}\ \bibnamefont
  {Blunden}}, \bibinfo {author} {\bibfnamefont {W.}~\bibnamefont
  {Melnitchouk}}, \ and\ \bibinfo {author} {\bibfnamefont {J.~A.}\ \bibnamefont
  {Tjon}},\ }\href {\doibase 10.1103/PhysRevLett.91.142304} {\bibfield
  {journal} {\bibinfo  {journal} {Phys. Rev. Lett.}\ }\textbf {\bibinfo
  {volume} {91}},\ \bibinfo {pages} {142304} (\bibinfo {year} {2003})},\
  \Eprint {http://arxiv.org/abs/nucl-th/0306076} {arXiv:nucl-th/0306076}
  \BibitemShut {NoStop}%
\bibitem [{\citenamefont {Guichon}\ and\ \citenamefont
  {Vanderhaeghen}(2003)}]{Guichon:2003qm}%
  \BibitemOpen
  \bibfield  {author} {\bibinfo {author} {\bibfnamefont {P.~A.~M.}\
  \bibnamefont {Guichon}}\ and\ \bibinfo {author} {\bibfnamefont
  {M.}~\bibnamefont {Vanderhaeghen}},\ }\href {\doibase
  10.1103/PhysRevLett.91.142303} {\bibfield  {journal} {\bibinfo  {journal}
  {Phys. Rev. Lett.}\ }\textbf {\bibinfo {volume} {91}},\ \bibinfo {pages}
  {142303} (\bibinfo {year} {2003})},\ \Eprint
  {http://arxiv.org/abs/hep-ph/0306007} {arXiv:hep-ph/0306007} \BibitemShut
  {NoStop}%
\bibitem [{\citenamefont {Afanasev}\ \emph {et~al.}(2017)\citenamefont
  {Afanasev}, \citenamefont {Blunden}, \citenamefont {Hasell},\ and\
  \citenamefont {Raue}}]{Afanasev:2017gsk}%
  \BibitemOpen
  \bibfield  {author} {\bibinfo {author} {\bibfnamefont {A.}~\bibnamefont
  {Afanasev}}, \bibinfo {author} {\bibfnamefont {P.~G.}\ \bibnamefont
  {Blunden}}, \bibinfo {author} {\bibfnamefont {D.}~\bibnamefont {Hasell}}, \
  and\ \bibinfo {author} {\bibfnamefont {B.~A.}\ \bibnamefont {Raue}},\ }\href
  {\doibase 10.1016/j.ppnp.2017.03.004} {\bibfield  {journal} {\bibinfo
  {journal} {Prog. Part. Nucl. Phys.}\ }\textbf {\bibinfo {volume} {95}},\
  \bibinfo {pages} {245} (\bibinfo {year} {2017})},\ \Eprint
  {http://arxiv.org/abs/1703.03874} {arXiv:1703.03874 [nucl-ex]} \BibitemShut
  {NoStop}%
\bibitem [{\citenamefont {Puckett}\ \emph {et~al.}(2017)\citenamefont {Puckett}
  \emph {et~al.}}]{Puckett:2017flj}%
  \BibitemOpen
  \bibfield  {author} {\bibinfo {author} {\bibfnamefont {A.~J.~R.}\
  \bibnamefont {Puckett}} \emph {et~al.},\ }\href {\doibase
  10.1103/PhysRevC.96.055203} {\bibfield  {journal} {\bibinfo  {journal} {Phys.
  Rev. C}\ }\textbf {\bibinfo {volume} {96}},\ \bibinfo {pages} {055203}
  (\bibinfo {year} {2017})},\ \bibinfo {note} {[Erratum: Phys.Rev.C 98, 019907
  (2018)]},\ \Eprint {http://arxiv.org/abs/1707.08587} {arXiv:1707.08587
  [nucl-ex]} \BibitemShut {NoStop}%
\bibitem [{\citenamefont {Kondratyuk}\ \emph {et~al.}(2005)\citenamefont
  {Kondratyuk}, \citenamefont {Blunden}, \citenamefont {Melnitchouk},\ and\
  \citenamefont {Tjon}}]{Kondratyuk:2005kk}%
  \BibitemOpen
  \bibfield  {author} {\bibinfo {author} {\bibfnamefont {S.}~\bibnamefont
  {Kondratyuk}}, \bibinfo {author} {\bibfnamefont {P.~G.}\ \bibnamefont
  {Blunden}}, \bibinfo {author} {\bibfnamefont {W.}~\bibnamefont
  {Melnitchouk}}, \ and\ \bibinfo {author} {\bibfnamefont {J.~A.}\ \bibnamefont
  {Tjon}},\ }\href {\doibase 10.1103/PhysRevLett.95.172503} {\bibfield
  {journal} {\bibinfo  {journal} {Phys. Rev. Lett.}\ }\textbf {\bibinfo
  {volume} {95}},\ \bibinfo {pages} {172503} (\bibinfo {year} {2005})},\
  \Eprint {http://arxiv.org/abs/nucl-th/0506026} {arXiv:nucl-th/0506026}
  \BibitemShut {NoStop}%
\bibitem [{\citenamefont {Blunden}\ and\ \citenamefont
  {Melnitchouk}(2017)}]{Blunden:2017nby}%
  \BibitemOpen
  \bibfield  {author} {\bibinfo {author} {\bibfnamefont {P.~G.}\ \bibnamefont
  {Blunden}}\ and\ \bibinfo {author} {\bibfnamefont {W.}~\bibnamefont
  {Melnitchouk}},\ }\href {\doibase 10.1103/PhysRevC.95.065209} {\bibfield
  {journal} {\bibinfo  {journal} {Phys. Rev. C}\ }\textbf {\bibinfo {volume}
  {95}},\ \bibinfo {pages} {065209} (\bibinfo {year} {2017})},\ \Eprint
  {http://arxiv.org/abs/1703.06181} {arXiv:1703.06181 [nucl-th]} \BibitemShut
  {NoStop}%
\bibitem [{\citenamefont {Kondratyuk}\ and\ \citenamefont
  {Blunden}(2007)}]{Kondratyuk:2007hc}%
  \BibitemOpen
  \bibfield  {author} {\bibinfo {author} {\bibfnamefont {S.}~\bibnamefont
  {Kondratyuk}}\ and\ \bibinfo {author} {\bibfnamefont {P.~G.}\ \bibnamefont
  {Blunden}},\ }\href {\doibase 10.1103/PhysRevC.75.038201} {\bibfield
  {journal} {\bibinfo  {journal} {Phys. Rev. C}\ }\textbf {\bibinfo {volume}
  {75}},\ \bibinfo {pages} {038201} (\bibinfo {year} {2007})},\ \Eprint
  {http://arxiv.org/abs/nucl-th/0701003} {arXiv:nucl-th/0701003} \BibitemShut
  {NoStop}%
\bibitem [{\citenamefont {Zhou}\ \emph {et~al.}(2007)\citenamefont {Zhou},
  \citenamefont {Kao},\ and\ \citenamefont {Yang}}]{Zhou:2007hr}%
  \BibitemOpen
  \bibfield  {author} {\bibinfo {author} {\bibfnamefont {H.~Q.}\ \bibnamefont
  {Zhou}}, \bibinfo {author} {\bibfnamefont {C.~W.}\ \bibnamefont {Kao}}, \
  and\ \bibinfo {author} {\bibfnamefont {S.~N.}\ \bibnamefont {Yang}},\ }\href
  {\doibase 10.1103/PhysRevLett.99.262001} {\bibfield  {journal} {\bibinfo
  {journal} {Phys. Rev. Lett.}\ }\textbf {\bibinfo {volume} {99}},\ \bibinfo
  {pages} {262001} (\bibinfo {year} {2007})},\ \bibinfo {note} {[Erratum:
  Phys.Rev.Lett. 100, 059903 (2008)]},\ \Eprint
  {http://arxiv.org/abs/0708.4297} {arXiv:0708.4297 [hep-ph]} \BibitemShut
  {NoStop}%
\bibitem [{\citenamefont {Tereshchuk}\ and\ \citenamefont
  {Afanasev}(2026)}]{Tereshchuk:2026iez}%
  \BibitemOpen
  \bibfield  {author} {\bibinfo {author} {\bibfnamefont {V.}~\bibnamefont
  {Tereshchuk}}\ and\ \bibinfo {author} {\bibfnamefont {A.}~\bibnamefont
  {Afanasev}},\ }\href@noop {} {\  (\bibinfo {year} {2026})},\ \Eprint
  {http://arxiv.org/abs/2607.20242} {arXiv:2607.20242 [nucl-th]} \BibitemShut
  {NoStop}%
\bibitem [{\citenamefont {Klest}(2026)}]{Klest:2025yik}%
  \BibitemOpen
  \bibfield  {author} {\bibinfo {author} {\bibfnamefont {H.~T.}\ \bibnamefont
  {Klest}},\ }\href {\doibase 10.1103/v6jm-jdxy} {\bibfield  {journal}
  {\bibinfo  {journal} {Phys. Rev. D}\ }\textbf {\bibinfo {volume} {113}},\
  \bibinfo {pages} {014045} (\bibinfo {year} {2026})},\ \Eprint
  {http://arxiv.org/abs/2507.23092} {arXiv:2507.23092 [hep-ph]} \BibitemShut
  {NoStop}%
\bibitem [{\citenamefont {Lee}\ and\ \citenamefont
  {Afanasev}(2025)}]{Lee:2025dsz}%
  \BibitemOpen
  \bibfield  {author} {\bibinfo {author} {\bibfnamefont {S.}~\bibnamefont
  {Lee}}\ and\ \bibinfo {author} {\bibfnamefont {A.}~\bibnamefont {Afanasev}},\
  }\href {\doibase 10.1103/4pwj-vp5p} {\bibfield  {journal} {\bibinfo
  {journal} {Phys. Rev. D}\ }\textbf {\bibinfo {volume} {111}},\ \bibinfo
  {pages} {113008} (\bibinfo {year} {2025})},\ \Eprint
  {http://arxiv.org/abs/2504.17123} {arXiv:2504.17123 [hep-ph]} \BibitemShut
  {NoStop}%
\bibitem [{\citenamefont {Fu}\ \emph {et~al.}(2022)\citenamefont {Fu},
  \citenamefont {Feng}, \citenamefont {Jin},\ and\ \citenamefont
  {Lu}}]{Fu:2022fgh}%
  \BibitemOpen
  \bibfield  {author} {\bibinfo {author} {\bibfnamefont {Y.}~\bibnamefont
  {Fu}}, \bibinfo {author} {\bibfnamefont {X.}~\bibnamefont {Feng}}, \bibinfo
  {author} {\bibfnamefont {L.-C.}\ \bibnamefont {Jin}}, \ and\ \bibinfo
  {author} {\bibfnamefont {C.-F.}\ \bibnamefont {Lu}},\ }\href {\doibase
  10.1103/PhysRevLett.128.172002} {\bibfield  {journal} {\bibinfo  {journal}
  {Phys. Rev. Lett.}\ }\textbf {\bibinfo {volume} {128}},\ \bibinfo {pages}
  {172002} (\bibinfo {year} {2022})},\ \Eprint
  {http://arxiv.org/abs/2202.01472} {arXiv:2202.01472 [hep-lat]} \BibitemShut
  {NoStop}%
\bibitem [{\citenamefont {Huang}\ \emph {et~al.}(2025)\citenamefont {Huang},
  \citenamefont {Shi}, \citenamefont {Wang},\ and\ \citenamefont
  {Zhao}}]{Huang:2024ugd}%
  \BibitemOpen
  \bibfield  {author} {\bibinfo {author} {\bibfnamefont {Y.-K.}\ \bibnamefont
  {Huang}}, \bibinfo {author} {\bibfnamefont {B.-X.}\ \bibnamefont {Shi}},
  \bibinfo {author} {\bibfnamefont {Y.-M.}\ \bibnamefont {Wang}}, \ and\
  \bibinfo {author} {\bibfnamefont {X.-C.}\ \bibnamefont {Zhao}},\ }\href
  {\doibase 10.1103/xz46-8ptk} {\bibfield  {journal} {\bibinfo  {journal}
  {Phys. Rev. Lett.}\ }\textbf {\bibinfo {volume} {135}},\ \bibinfo {pages}
  {061901} (\bibinfo {year} {2025})},\ \Eprint
  {http://arxiv.org/abs/2407.18724} {arXiv:2407.18724 [hep-ph]} \BibitemShut
  {NoStop}%
\bibitem [{\citenamefont {Chen}\ \emph
  {et~al.}(2025{\natexlab{a}})\citenamefont {Chen}, \citenamefont {Chen},
  \citenamefont {Feng}, \citenamefont {Hu},\ and\ \citenamefont
  {Jia}}]{Chen:2024fhj}%
  \BibitemOpen
  \bibfield  {author} {\bibinfo {author} {\bibfnamefont {L.-B.}\ \bibnamefont
  {Chen}}, \bibinfo {author} {\bibfnamefont {W.}~\bibnamefont {Chen}}, \bibinfo
  {author} {\bibfnamefont {F.}~\bibnamefont {Feng}}, \bibinfo {author}
  {\bibfnamefont {S.}~\bibnamefont {Hu}}, \ and\ \bibinfo {author}
  {\bibfnamefont {Y.}~\bibnamefont {Jia}},\ }\href {\doibase 10.1103/r25y-rsz5}
  {\bibfield  {journal} {\bibinfo  {journal} {Phys. Rev. Lett.}\ }\textbf
  {\bibinfo {volume} {135}},\ \bibinfo {pages} {131903} (\bibinfo {year}
  {2025}{\natexlab{a}})},\ \Eprint {http://arxiv.org/abs/2406.19994}
  {arXiv:2406.19994 [hep-ph]} \BibitemShut {NoStop}%
\bibitem [{\citenamefont {Yu}\ \emph {et~al.}(2026)\citenamefont {Yu},
  \citenamefont {Cheng}, \citenamefont {Han}, \citenamefont {Li},\ and\
  \citenamefont {Yu}}]{Yu:2025jcs}%
  \BibitemOpen
  \bibfield  {author} {\bibinfo {author} {\bibfnamefont {J.-X.}\ \bibnamefont
  {Yu}}, \bibinfo {author} {\bibfnamefont {S.}~\bibnamefont {Cheng}}, \bibinfo
  {author} {\bibfnamefont {J.-J.}\ \bibnamefont {Han}}, \bibinfo {author}
  {\bibfnamefont {H.-n.}\ \bibnamefont {Li}}, \ and\ \bibinfo {author}
  {\bibfnamefont {F.-S.}\ \bibnamefont {Yu}},\ }\href {\doibase
  10.1140/epjc/s10052-026-15554-x} {\bibfield  {journal} {\bibinfo  {journal}
  {Eur. Phys. J. C}\ }\textbf {\bibinfo {volume} {86}},\ \bibinfo {pages} {300}
  (\bibinfo {year} {2026})},\ \Eprint {http://arxiv.org/abs/2511.12589}
  {arXiv:2511.12589 [hep-ph]} \BibitemShut {NoStop}%
\bibitem [{\citenamefont {Lin}\ and\ \citenamefont
  {Orginos}(2009)}]{Lin:2008mr}%
  \BibitemOpen
  \bibfield  {author} {\bibinfo {author} {\bibfnamefont {H.-W.}\ \bibnamefont
  {Lin}}\ and\ \bibinfo {author} {\bibfnamefont {K.}~\bibnamefont {Orginos}},\
  }\href {\doibase 10.1103/PhysRevD.79.074507} {\bibfield  {journal} {\bibinfo
  {journal} {Phys. Rev. D}\ }\textbf {\bibinfo {volume} {79}},\ \bibinfo
  {pages} {074507} (\bibinfo {year} {2009})},\ \Eprint
  {http://arxiv.org/abs/0812.4456} {arXiv:0812.4456 [hep-lat]} \BibitemShut
  {NoStop}%
\bibitem [{\citenamefont {Shanahan}\ \emph
  {et~al.}(2014{\natexlab{a}})\citenamefont {Shanahan}, \citenamefont {Thomas},
  \citenamefont {Young}, \citenamefont {Zanotti}, \citenamefont {Horsley},
  \citenamefont {Nakamura}, \citenamefont {Pleiter}, \citenamefont {Rakow},
  \citenamefont {Schierholz},\ and\ \citenamefont
  {St{\"u}ben}}]{Shanahan:2014cga}%
  \BibitemOpen
  \bibfield  {author} {\bibinfo {author} {\bibfnamefont {P.~E.}\ \bibnamefont
  {Shanahan}}, \bibinfo {author} {\bibfnamefont {A.~W.}\ \bibnamefont
  {Thomas}}, \bibinfo {author} {\bibfnamefont {R.~D.}\ \bibnamefont {Young}},
  \bibinfo {author} {\bibfnamefont {J.~M.}\ \bibnamefont {Zanotti}}, \bibinfo
  {author} {\bibfnamefont {R.}~\bibnamefont {Horsley}}, \bibinfo {author}
  {\bibfnamefont {Y.}~\bibnamefont {Nakamura}}, \bibinfo {author}
  {\bibfnamefont {D.}~\bibnamefont {Pleiter}}, \bibinfo {author} {\bibfnamefont
  {P.~E.~L.}\ \bibnamefont {Rakow}}, \bibinfo {author} {\bibfnamefont
  {G.}~\bibnamefont {Schierholz}}, \ and\ \bibinfo {author} {\bibfnamefont
  {H.}~\bibnamefont {St{\"u}ben}},\ }\href {\doibase
  10.1103/PhysRevD.90.034502} {\bibfield  {journal} {\bibinfo  {journal} {Phys.
  Rev. D}\ }\textbf {\bibinfo {volume} {90}},\ \bibinfo {pages} {034502}
  (\bibinfo {year} {2014}{\natexlab{a}})},\ \Eprint
  {http://arxiv.org/abs/1403.1965} {arXiv:1403.1965 [hep-lat]} \BibitemShut
  {NoStop}%
\bibitem [{\citenamefont {Shanahan}\ \emph
  {et~al.}(2014{\natexlab{b}})\citenamefont {Shanahan}, \citenamefont {Thomas},
  \citenamefont {Young}, \citenamefont {Zanotti}, \citenamefont {Horsley},
  \citenamefont {Nakamura}, \citenamefont {Pleiter}, \citenamefont {Rakow},
  \citenamefont {Schierholz},\ and\ \citenamefont {St{\"u}ben}}]{CSSM:2014knt}%
  \BibitemOpen
  \bibfield  {author} {\bibinfo {author} {\bibfnamefont {P.~E.}\ \bibnamefont
  {Shanahan}}, \bibinfo {author} {\bibfnamefont {A.~W.}\ \bibnamefont
  {Thomas}}, \bibinfo {author} {\bibfnamefont {R.~D.}\ \bibnamefont {Young}},
  \bibinfo {author} {\bibfnamefont {J.~M.}\ \bibnamefont {Zanotti}}, \bibinfo
  {author} {\bibfnamefont {R.}~\bibnamefont {Horsley}}, \bibinfo {author}
  {\bibfnamefont {Y.}~\bibnamefont {Nakamura}}, \bibinfo {author}
  {\bibfnamefont {D.}~\bibnamefont {Pleiter}}, \bibinfo {author} {\bibfnamefont
  {P.~E.~L.}\ \bibnamefont {Rakow}}, \bibinfo {author} {\bibfnamefont
  {G.}~\bibnamefont {Schierholz}}, \ and\ \bibinfo {author} {\bibfnamefont
  {H.}~\bibnamefont {St{\"u}ben}} (\bibinfo {collaboration} {CSSM,
  QCDSF/UKQCD}),\ }\href {\doibase 10.1103/PhysRevD.89.074511} {\bibfield
  {journal} {\bibinfo  {journal} {Phys. Rev. D}\ }\textbf {\bibinfo {volume}
  {89}},\ \bibinfo {pages} {074511} (\bibinfo {year} {2014}{\natexlab{b}})},\
  \Eprint {http://arxiv.org/abs/1401.5862} {arXiv:1401.5862 [hep-lat]}
  \BibitemShut {NoStop}%
\bibitem [{\citenamefont {Jang}\ \emph {et~al.}(2018)\citenamefont {Jang},
  \citenamefont {Bhattacharya}, \citenamefont {Gupta}, \citenamefont {Lin},\
  and\ \citenamefont {Yoon}}]{Jang:2018djx}%
  \BibitemOpen
  \bibfield  {author} {\bibinfo {author} {\bibfnamefont {Y.-C.}\ \bibnamefont
  {Jang}}, \bibinfo {author} {\bibfnamefont {T.}~\bibnamefont {Bhattacharya}},
  \bibinfo {author} {\bibfnamefont {R.}~\bibnamefont {Gupta}}, \bibinfo
  {author} {\bibfnamefont {H.-W.}\ \bibnamefont {Lin}}, \ and\ \bibinfo
  {author} {\bibfnamefont {B.}~\bibnamefont {Yoon}} (\bibinfo {collaboration}
  {PNDME}),\ }\href {\doibase 10.22323/1.334.0123} {\bibfield  {journal}
  {\bibinfo  {journal} {PoS}\ }\textbf {\bibinfo {volume} {LATTICE2018}},\
  \bibinfo {pages} {123} (\bibinfo {year} {2018})},\ \Eprint
  {http://arxiv.org/abs/1901.00060} {arXiv:1901.00060 [hep-lat]} \BibitemShut
  {NoStop}%
\bibitem [{\citenamefont {Lin}(2021)}]{Lin:2020rxa}%
  \BibitemOpen
  \bibfield  {author} {\bibinfo {author} {\bibfnamefont {H.-W.}\ \bibnamefont
  {Lin}},\ }\href {\doibase 10.1103/PhysRevLett.127.182001} {\bibfield
  {journal} {\bibinfo  {journal} {Phys. Rev. Lett.}\ }\textbf {\bibinfo
  {volume} {127}},\ \bibinfo {pages} {182001} (\bibinfo {year} {2021})},\
  \Eprint {http://arxiv.org/abs/2008.12474} {arXiv:2008.12474 [hep-ph]}
  \BibitemShut {NoStop}%
\bibitem [{\citenamefont {Constantinou}\ \emph {et~al.}(2021)\citenamefont
  {Constantinou} \emph {et~al.}}]{Constantinou:2020hdm}%
  \BibitemOpen
  \bibfield  {author} {\bibinfo {author} {\bibfnamefont {M.}~\bibnamefont
  {Constantinou}} \emph {et~al.},\ }\href {\doibase 10.1016/j.ppnp.2021.103908}
  {\bibfield  {journal} {\bibinfo  {journal} {Prog. Part. Nucl. Phys.}\
  }\textbf {\bibinfo {volume} {121}},\ \bibinfo {pages} {103908} (\bibinfo
  {year} {2021})},\ \Eprint {http://arxiv.org/abs/2006.08636} {arXiv:2006.08636
  [hep-ph]} \BibitemShut {NoStop}%
\bibitem [{\citenamefont {Liu}\ and\ \citenamefont {Huang}(2009)}]{Liu:2009mb}%
  \BibitemOpen
  \bibfield  {author} {\bibinfo {author} {\bibfnamefont {Y.-L.}\ \bibnamefont
  {Liu}}\ and\ \bibinfo {author} {\bibfnamefont {M.-Q.}\ \bibnamefont
  {Huang}},\ }\href {\doibase 10.1103/PhysRevD.79.114031} {\bibfield  {journal}
  {\bibinfo  {journal} {Phys. Rev. D}\ }\textbf {\bibinfo {volume} {79}},\
  \bibinfo {pages} {114031} (\bibinfo {year} {2009})},\ \Eprint
  {http://arxiv.org/abs/0906.1935} {arXiv:0906.1935 [hep-ph]} \BibitemShut
  {NoStop}%
\bibitem [{\citenamefont {Ramalho}\ \emph {et~al.}(2020)\citenamefont
  {Ramalho}, \citenamefont {Pe{\~n}a},\ and\ \citenamefont
  {Tsushima}}]{Ramalho:2019koj}%
  \BibitemOpen
  \bibfield  {author} {\bibinfo {author} {\bibfnamefont {G.}~\bibnamefont
  {Ramalho}}, \bibinfo {author} {\bibfnamefont {M.~T.}\ \bibnamefont
  {Pe{\~n}a}}, \ and\ \bibinfo {author} {\bibfnamefont {K.}~\bibnamefont
  {Tsushima}},\ }\href {\doibase 10.1103/PhysRevD.101.014014} {\bibfield
  {journal} {\bibinfo  {journal} {Phys. Rev. D}\ }\textbf {\bibinfo {volume}
  {101}},\ \bibinfo {pages} {014014} (\bibinfo {year} {2020})},\ \Eprint
  {http://arxiv.org/abs/1908.04864} {arXiv:1908.04864 [hep-ph]} \BibitemShut
  {NoStop}%
\bibitem [{\citenamefont {Ramalho}\ and\ \citenamefont
  {Tsushima}(2012)}]{Ramalho:2012ad}%
  \BibitemOpen
  \bibfield  {author} {\bibinfo {author} {\bibfnamefont {G.}~\bibnamefont
  {Ramalho}}\ and\ \bibinfo {author} {\bibfnamefont {K.}~\bibnamefont
  {Tsushima}},\ }\href {\doibase 10.1103/PhysRevD.86.114030} {\bibfield
  {journal} {\bibinfo  {journal} {Phys. Rev. D}\ }\textbf {\bibinfo {volume}
  {86}},\ \bibinfo {pages} {114030} (\bibinfo {year} {2012})},\ \Eprint
  {http://arxiv.org/abs/1210.7465} {arXiv:1210.7465 [hep-ph]} \BibitemShut
  {NoStop}%
\bibitem [{\citenamefont {Kubis}\ and\ \citenamefont
  {Meissner}(2001)}]{Kubis:2000aa}%
  \BibitemOpen
  \bibfield  {author} {\bibinfo {author} {\bibfnamefont {B.}~\bibnamefont
  {Kubis}}\ and\ \bibinfo {author} {\bibfnamefont {U.~G.}\ \bibnamefont
  {Meissner}},\ }\href {\doibase 10.1007/s100520100570} {\bibfield  {journal}
  {\bibinfo  {journal} {Eur. Phys. J. C}\ }\textbf {\bibinfo {volume} {18}},\
  \bibinfo {pages} {747} (\bibinfo {year} {2001})},\ \Eprint
  {http://arxiv.org/abs/hep-ph/0010283} {arXiv:hep-ph/0010283} \BibitemShut
  {NoStop}%
\bibitem [{\citenamefont {Sanchis-Alepuz}\ \emph {et~al.}(2018)\citenamefont
  {Sanchis-Alepuz}, \citenamefont {Alkofer},\ and\ \citenamefont
  {Fischer}}]{Sanchis-Alepuz:2017mir}%
  \BibitemOpen
  \bibfield  {author} {\bibinfo {author} {\bibfnamefont {H.}~\bibnamefont
  {Sanchis-Alepuz}}, \bibinfo {author} {\bibfnamefont {R.}~\bibnamefont
  {Alkofer}}, \ and\ \bibinfo {author} {\bibfnamefont {C.~S.}\ \bibnamefont
  {Fischer}},\ }\href {\doibase 10.1140/epja/i2018-12465-x} {\bibfield
  {journal} {\bibinfo  {journal} {Eur. Phys. J. A}\ }\textbf {\bibinfo {volume}
  {54}},\ \bibinfo {pages} {41} (\bibinfo {year} {2018})},\ \Eprint
  {http://arxiv.org/abs/1707.08463} {arXiv:1707.08463 [hep-ph]} \BibitemShut
  {NoStop}%
\bibitem [{\citenamefont {Sanchis-Alepuz}\ and\ \citenamefont
  {Fischer}(2016)}]{Sanchis-Alepuz:2015fcg}%
  \BibitemOpen
  \bibfield  {author} {\bibinfo {author} {\bibfnamefont {H.}~\bibnamefont
  {Sanchis-Alepuz}}\ and\ \bibinfo {author} {\bibfnamefont {C.~S.}\
  \bibnamefont {Fischer}},\ }\href {\doibase 10.1140/epja/i2016-16034-1}
  {\bibfield  {journal} {\bibinfo  {journal} {Eur. Phys. J. A}\ }\textbf
  {\bibinfo {volume} {52}},\ \bibinfo {pages} {34} (\bibinfo {year} {2016})},\
  \Eprint {http://arxiv.org/abs/1512.00833} {arXiv:1512.00833 [hep-ph]}
  \BibitemShut {NoStop}%
\bibitem [{\citenamefont {Liu}\ and\ \citenamefont
  {Fischer}(2024)}]{Liu:2023reo}%
  \BibitemOpen
  \bibfield  {author} {\bibinfo {author} {\bibfnamefont {L.}~\bibnamefont
  {Liu}}\ and\ \bibinfo {author} {\bibfnamefont {C.~S.}\ \bibnamefont
  {Fischer}},\ }\href {\doibase 10.1140/epja/s10050-024-01283-w} {\bibfield
  {journal} {\bibinfo  {journal} {Eur. Phys. J. A}\ }\textbf {\bibinfo {volume}
  {60}},\ \bibinfo {pages} {84} (\bibinfo {year} {2024})},\ \Eprint
  {http://arxiv.org/abs/2311.13269} {arXiv:2311.13269 [hep-ph]} \BibitemShut
  {NoStop}%
\bibitem [{\citenamefont {Cheng}\ \emph {et~al.}(2025)\citenamefont {Cheng},
  \citenamefont {Yao}, \citenamefont {Binosi}, \citenamefont {Lu},\ and\
  \citenamefont {Roberts}}]{Cheng:2025yij}%
  \BibitemOpen
  \bibfield  {author} {\bibinfo {author} {\bibfnamefont {P.}~\bibnamefont
  {Cheng}}, \bibinfo {author} {\bibfnamefont {Z.~Q.}\ \bibnamefont {Yao}},
  \bibinfo {author} {\bibfnamefont {D.}~\bibnamefont {Binosi}}, \bibinfo
  {author} {\bibfnamefont {Y.}~\bibnamefont {Lu}}, \ and\ \bibinfo {author}
  {\bibfnamefont {C.~D.}\ \bibnamefont {Roberts}},\ }\href {\doibase
  10.1140/epja/s10050-025-01724-0} {\bibfield  {journal} {\bibinfo  {journal}
  {Eur. Phys. J. A}\ }\textbf {\bibinfo {volume} {61}},\ \bibinfo {pages} {255}
  (\bibinfo {year} {2025})},\ \Eprint {http://arxiv.org/abs/2507.13484}
  {arXiv:2507.13484 [hep-ph]} \BibitemShut {NoStop}%
\bibitem [{\citenamefont {Rekalo}\ and\ \citenamefont
  {Tomasi-Gustafsson}(2004{\natexlab{a}})}]{Rekalo:2003xa}%
  \BibitemOpen
  \bibfield  {author} {\bibinfo {author} {\bibfnamefont {M.~P.}\ \bibnamefont
  {Rekalo}}\ and\ \bibinfo {author} {\bibfnamefont {E.}~\bibnamefont
  {Tomasi-Gustafsson}},\ }\href {\doibase 10.1140/epja/i2004-10039-3}
  {\bibfield  {journal} {\bibinfo  {journal} {Eur. Phys. J. A}\ }\textbf
  {\bibinfo {volume} {22}},\ \bibinfo {pages} {331} (\bibinfo {year}
  {2004}{\natexlab{a}})},\ \Eprint {http://arxiv.org/abs/nucl-th/0307066}
  {arXiv:nucl-th/0307066} \BibitemShut {NoStop}%
\bibitem [{\citenamefont {Rekalo}\ and\ \citenamefont
  {Tomasi-Gustafsson}(2004{\natexlab{b}})}]{Rekalo:2003km}%
  \BibitemOpen
  \bibfield  {author} {\bibinfo {author} {\bibfnamefont {M.~P.}\ \bibnamefont
  {Rekalo}}\ and\ \bibinfo {author} {\bibfnamefont {E.}~\bibnamefont
  {Tomasi-Gustafsson}},\ }\href {\doibase 10.1016/j.nuclphysa.2004.04.111}
  {\bibfield  {journal} {\bibinfo  {journal} {Nucl. Phys. A}\ }\textbf
  {\bibinfo {volume} {740}},\ \bibinfo {pages} {271} (\bibinfo {year}
  {2004}{\natexlab{b}})},\ \Eprint {http://arxiv.org/abs/nucl-th/0312030}
  {arXiv:nucl-th/0312030} \BibitemShut {NoStop}%
\bibitem [{\citenamefont {Rekalo}\ and\ \citenamefont
  {Tomasi-Gustafsson}(2004{\natexlab{c}})}]{Rekalo:2004wa}%
  \BibitemOpen
  \bibfield  {author} {\bibinfo {author} {\bibfnamefont {M.~P.}\ \bibnamefont
  {Rekalo}}\ and\ \bibinfo {author} {\bibfnamefont {E.}~\bibnamefont
  {Tomasi-Gustafsson}},\ }\href {\doibase 10.1016/j.nuclphysa.2004.07.009}
  {\bibfield  {journal} {\bibinfo  {journal} {Nucl. Phys. A}\ }\textbf
  {\bibinfo {volume} {742}},\ \bibinfo {pages} {322} (\bibinfo {year}
  {2004}{\natexlab{c}})},\ \Eprint {http://arxiv.org/abs/nucl-th/0402004}
  {arXiv:nucl-th/0402004} \BibitemShut {NoStop}%
\bibitem [{\citenamefont {Gakh}\ and\ \citenamefont
  {Tomasi-Gustafsson}(2006)}]{Gakh:2005hh}%
  \BibitemOpen
  \bibfield  {author} {\bibinfo {author} {\bibfnamefont {G.~I.}\ \bibnamefont
  {Gakh}}\ and\ \bibinfo {author} {\bibfnamefont {E.}~\bibnamefont
  {Tomasi-Gustafsson}},\ }\href {\doibase 10.1016/j.nuclphysa.2006.03.009}
  {\bibfield  {journal} {\bibinfo  {journal} {Nucl. Phys. A}\ }\textbf
  {\bibinfo {volume} {771}},\ \bibinfo {pages} {169} (\bibinfo {year}
  {2006})},\ \Eprint {http://arxiv.org/abs/hep-ph/0511240}
  {arXiv:hep-ph/0511240} \BibitemShut {NoStop}%
\bibitem [{\citenamefont {Gakh}\ and\ \citenamefont
  {Tomasi-Gustafsson}(2005)}]{Gakh:2005wa}%
  \BibitemOpen
  \bibfield  {author} {\bibinfo {author} {\bibfnamefont {G.~I.}\ \bibnamefont
  {Gakh}}\ and\ \bibinfo {author} {\bibfnamefont {E.}~\bibnamefont
  {Tomasi-Gustafsson}},\ }\href {\doibase 10.1016/j.nuclphysa.2005.07.002}
  {\bibfield  {journal} {\bibinfo  {journal} {Nucl. Phys. A}\ }\textbf
  {\bibinfo {volume} {761}},\ \bibinfo {pages} {120} (\bibinfo {year}
  {2005})},\ \Eprint {http://arxiv.org/abs/nucl-th/0504021}
  {arXiv:nucl-th/0504021} \BibitemShut {NoStop}%
\bibitem [{\citenamefont {Tomasi-Gustafsson}\ \emph {et~al.}(2005)\citenamefont
  {Tomasi-Gustafsson}, \citenamefont {Lacroix}, \citenamefont {Duterte},\ and\
  \citenamefont {Gakh}}]{Tomasi-Gustafsson:2005svz}%
  \BibitemOpen
  \bibfield  {author} {\bibinfo {author} {\bibfnamefont {E.}~\bibnamefont
  {Tomasi-Gustafsson}}, \bibinfo {author} {\bibfnamefont {F.}~\bibnamefont
  {Lacroix}}, \bibinfo {author} {\bibfnamefont {C.}~\bibnamefont {Duterte}}, \
  and\ \bibinfo {author} {\bibfnamefont {G.~I.}\ \bibnamefont {Gakh}},\ }\href
  {\doibase 10.1140/epja/i2005-10030-6} {\bibfield  {journal} {\bibinfo
  {journal} {Eur. Phys. J. A}\ }\textbf {\bibinfo {volume} {24}},\ \bibinfo
  {pages} {419} (\bibinfo {year} {2005})},\ \Eprint
  {http://arxiv.org/abs/nucl-th/0503001} {arXiv:nucl-th/0503001} \BibitemShut
  {NoStop}%
\bibitem [{\citenamefont {Adamuscin}\ \emph {et~al.}(2007)\citenamefont
  {Adamuscin}, \citenamefont {Gakh},\ and\ \citenamefont
  {Tomasi-Gustafsson}}]{Adamuscin:2007xn}%
  \BibitemOpen
  \bibfield  {author} {\bibinfo {author} {\bibfnamefont {C.}~\bibnamefont
  {Adamuscin}}, \bibinfo {author} {\bibfnamefont {G.~I.}\ \bibnamefont {Gakh}},
  \ and\ \bibinfo {author} {\bibfnamefont {E.}~\bibnamefont
  {Tomasi-Gustafsson}},\ }\href@noop {} {\  (\bibinfo {year} {2007})},\ \Eprint
  {http://arxiv.org/abs/0704.3375} {arXiv:0704.3375 [hep-ph]} \BibitemShut
  {NoStop}%
\bibitem [{\citenamefont {Chen}\ \emph {et~al.}(2008)\citenamefont {Chen},
  \citenamefont {Zhou},\ and\ \citenamefont {Dong}}]{Chen:2008hka}%
  \BibitemOpen
  \bibfield  {author} {\bibinfo {author} {\bibfnamefont {D.~Y.}\ \bibnamefont
  {Chen}}, \bibinfo {author} {\bibfnamefont {H.~Q.}\ \bibnamefont {Zhou}}, \
  and\ \bibinfo {author} {\bibfnamefont {Y.~B.}\ \bibnamefont {Dong}},\ }\href
  {\doibase 10.1103/PhysRevC.78.045208} {\bibfield  {journal} {\bibinfo
  {journal} {Phys. Rev. C}\ }\textbf {\bibinfo {volume} {78}},\ \bibinfo
  {pages} {045208} (\bibinfo {year} {2008})},\ \Eprint
  {http://arxiv.org/abs/0806.2489} {arXiv:0806.2489 [nucl-th]} \BibitemShut
  {NoStop}%
\bibitem [{\citenamefont {Qian}\ \emph {et~al.}(2023)\citenamefont {Qian},
  \citenamefont {Liu}, \citenamefont {Cao},\ and\ \citenamefont
  {Liu}}]{Qian:2022whn}%
  \BibitemOpen
  \bibfield  {author} {\bibinfo {author} {\bibfnamefont {R.-Q.}\ \bibnamefont
  {Qian}}, \bibinfo {author} {\bibfnamefont {Z.-W.}\ \bibnamefont {Liu}},
  \bibinfo {author} {\bibfnamefont {X.}~\bibnamefont {Cao}}, \ and\ \bibinfo
  {author} {\bibfnamefont {X.}~\bibnamefont {Liu}},\ }\href {\doibase
  10.1103/PhysRevD.107.L091502} {\bibfield  {journal} {\bibinfo  {journal}
  {Phys. Rev. D}\ }\textbf {\bibinfo {volume} {107}},\ \bibinfo {pages}
  {L091502} (\bibinfo {year} {2023})},\ \Eprint
  {http://arxiv.org/abs/2211.11555} {arXiv:2211.11555 [nucl-th]} \BibitemShut
  {NoStop}%
\bibitem [{\citenamefont {Dai}\ \emph {et~al.}(2023)\citenamefont {Dai},
  \citenamefont {Cao},\ and\ \citenamefont {Lenske}}]{Dai:2023vsw}%
  \BibitemOpen
  \bibfield  {author} {\bibinfo {author} {\bibfnamefont {J.-P.}\ \bibnamefont
  {Dai}}, \bibinfo {author} {\bibfnamefont {X.}~\bibnamefont {Cao}}, \ and\
  \bibinfo {author} {\bibfnamefont {H.}~\bibnamefont {Lenske}},\ }\href
  {\doibase 10.1016/j.physletb.2023.138192} {\bibfield  {journal} {\bibinfo
  {journal} {Phys. Lett. B}\ }\textbf {\bibinfo {volume} {846}},\ \bibinfo
  {pages} {138192} (\bibinfo {year} {2023})},\ \Eprint
  {http://arxiv.org/abs/2304.04913} {arXiv:2304.04913 [hep-ph]} \BibitemShut
  {NoStop}%
\bibitem [{\citenamefont {Cao}\ \emph {et~al.}(2022)\citenamefont {Cao},
  \citenamefont {Dai},\ and\ \citenamefont {Lenske}}]{Cao:2021asd}%
  \BibitemOpen
  \bibfield  {author} {\bibinfo {author} {\bibfnamefont {X.}~\bibnamefont
  {Cao}}, \bibinfo {author} {\bibfnamefont {J.-P.}\ \bibnamefont {Dai}}, \ and\
  \bibinfo {author} {\bibfnamefont {H.}~\bibnamefont {Lenske}},\ }\href
  {\doibase 10.1103/PhysRevD.105.L071503} {\bibfield  {journal} {\bibinfo
  {journal} {Phys. Rev. D}\ }\textbf {\bibinfo {volume} {105}},\ \bibinfo
  {pages} {L071503} (\bibinfo {year} {2022})},\ \Eprint
  {http://arxiv.org/abs/2109.15132} {arXiv:2109.15132 [hep-ph]} \BibitemShut
  {NoStop}%
\bibitem [{\citenamefont {Xiao}\ \emph {et~al.}(2019)\citenamefont {Xiao},
  \citenamefont {Weng}, \citenamefont {Zhong},\ and\ \citenamefont
  {Zhu}}]{Xiao:2019qhl}%
  \BibitemOpen
  \bibfield  {author} {\bibinfo {author} {\bibfnamefont {L.-Y.}\ \bibnamefont
  {Xiao}}, \bibinfo {author} {\bibfnamefont {X.-Z.}\ \bibnamefont {Weng}},
  \bibinfo {author} {\bibfnamefont {X.-H.}\ \bibnamefont {Zhong}}, \ and\
  \bibinfo {author} {\bibfnamefont {S.-L.}\ \bibnamefont {Zhu}},\ }\href
  {\doibase 10.1088/1674-1137/43/11/113105} {\bibfield  {journal} {\bibinfo
  {journal} {Chin. Phys. C}\ }\textbf {\bibinfo {volume} {43}},\ \bibinfo
  {pages} {113105} (\bibinfo {year} {2019})},\ \Eprint
  {http://arxiv.org/abs/1904.06616} {arXiv:1904.06616 [hep-ph]} \BibitemShut
  {NoStop}%
\bibitem [{\citenamefont {Yan}\ \emph {et~al.}(2024)\citenamefont {Yan},
  \citenamefont {Chen}, \citenamefont {Li},\ and\ \citenamefont
  {Xie}}]{Yan:2023nlb}%
  \BibitemOpen
  \bibfield  {author} {\bibinfo {author} {\bibfnamefont {B.}~\bibnamefont
  {Yan}}, \bibinfo {author} {\bibfnamefont {C.}~\bibnamefont {Chen}}, \bibinfo
  {author} {\bibfnamefont {X.}~\bibnamefont {Li}}, \ and\ \bibinfo {author}
  {\bibfnamefont {J.-J.}\ \bibnamefont {Xie}},\ }\href {\doibase
  10.1103/PhysRevD.109.036033} {\bibfield  {journal} {\bibinfo  {journal}
  {Phys. Rev. D}\ }\textbf {\bibinfo {volume} {109}},\ \bibinfo {pages}
  {036033} (\bibinfo {year} {2024})},\ \Eprint
  {http://arxiv.org/abs/2312.04866} {arXiv:2312.04866 [nucl-th]} \BibitemShut
  {NoStop}%
\bibitem [{\citenamefont {Haidenbauer}\ \emph {et~al.}(2021)\citenamefont
  {Haidenbauer}, \citenamefont {Mei{\ss}ner},\ and\ \citenamefont
  {Dai}}]{Haidenbauer:2020wyp}%
  \BibitemOpen
  \bibfield  {author} {\bibinfo {author} {\bibfnamefont {J.}~\bibnamefont
  {Haidenbauer}}, \bibinfo {author} {\bibfnamefont {U.-G.}\ \bibnamefont
  {Mei{\ss}ner}}, \ and\ \bibinfo {author} {\bibfnamefont {L.-Y.}\ \bibnamefont
  {Dai}},\ }\href {\doibase 10.1103/PhysRevD.103.014028} {\bibfield  {journal}
  {\bibinfo  {journal} {Phys. Rev. D}\ }\textbf {\bibinfo {volume} {103}},\
  \bibinfo {pages} {014028} (\bibinfo {year} {2021})},\ \Eprint
  {http://arxiv.org/abs/2011.06857} {arXiv:2011.06857 [nucl-th]} \BibitemShut
  {NoStop}%
\bibitem [{\citenamefont {Dai}\ \emph {et~al.}(2025)\citenamefont {Dai},
  \citenamefont {Haidenbauer},\ and\ \citenamefont
  {Mei{\ss}ner}}]{Dai:2024lau}%
  \BibitemOpen
  \bibfield  {author} {\bibinfo {author} {\bibfnamefont {L.-Y.}\ \bibnamefont
  {Dai}}, \bibinfo {author} {\bibfnamefont {J.}~\bibnamefont {Haidenbauer}}, \
  and\ \bibinfo {author} {\bibfnamefont {U.-G.}\ \bibnamefont {Mei{\ss}ner}},\
  }\href {\doibase 10.1088/0256-307X/42/3/030202} {\bibfield  {journal}
  {\bibinfo  {journal} {Chin. Phys. Lett.}\ }\textbf {\bibinfo {volume} {42}},\
  \bibinfo {pages} {030202} (\bibinfo {year} {2025})},\ \Eprint
  {http://arxiv.org/abs/2412.07543} {arXiv:2412.07543 [hep-ph]} \BibitemShut
  {NoStop}%
\bibitem [{\citenamefont {Jia}\ \emph {et~al.}(2025)\citenamefont {Jia},
  \citenamefont {Zhang}, \citenamefont {Guo},\ and\ \citenamefont
  {Li}}]{Jia:2024ybo}%
  \BibitemOpen
  \bibfield  {author} {\bibinfo {author} {\bibfnamefont {Z.-S.}\ \bibnamefont
  {Jia}}, \bibinfo {author} {\bibfnamefont {Z.-H.}\ \bibnamefont {Zhang}},
  \bibinfo {author} {\bibfnamefont {F.-K.}\ \bibnamefont {Guo}}, \ and\
  \bibinfo {author} {\bibfnamefont {G.}~\bibnamefont {Li}},\ }\href {\doibase
  10.1103/PhysRevD.111.054014} {\bibfield  {journal} {\bibinfo  {journal}
  {Phys. Rev. D}\ }\textbf {\bibinfo {volume} {111}},\ \bibinfo {pages}
  {054014} (\bibinfo {year} {2025})},\ \Eprint
  {http://arxiv.org/abs/2410.16873} {arXiv:2410.16873 [hep-ph]} \BibitemShut
  {NoStop}%
\bibitem [{\citenamefont {Yang}\ \emph {et~al.}(2024)\citenamefont {Yang},
  \citenamefont {Guo}, \citenamefont {Li}, \citenamefont {Dai}, \citenamefont
  {Haidenbauer},\ and\ \citenamefont {Mei{\ss}ner}}]{Yang:2024iuc}%
  \BibitemOpen
  \bibfield  {author} {\bibinfo {author} {\bibfnamefont {Q.-H.}\ \bibnamefont
  {Yang}}, \bibinfo {author} {\bibfnamefont {D.}~\bibnamefont {Guo}}, \bibinfo
  {author} {\bibfnamefont {M.-Y.}\ \bibnamefont {Li}}, \bibinfo {author}
  {\bibfnamefont {L.-Y.}\ \bibnamefont {Dai}}, \bibinfo {author} {\bibfnamefont
  {J.}~\bibnamefont {Haidenbauer}}, \ and\ \bibinfo {author} {\bibfnamefont
  {U.-G.}\ \bibnamefont {Mei{\ss}ner}},\ }\href {\doibase
  10.1007/JHEP08(2024)208} {\bibfield  {journal} {\bibinfo  {journal} {JHEP}\
  }\textbf {\bibinfo {volume} {08}},\ \bibinfo {pages} {208} (\bibinfo {year}
  {2024})},\ \Eprint {http://arxiv.org/abs/2404.12448} {arXiv:2404.12448
  [nucl-th]} \BibitemShut {NoStop}%
\bibitem [{\citenamefont {Bianconi}\ and\ \citenamefont
  {Tomasi-Gustafsson}(2015)}]{Bianconi:2015owa}%
  \BibitemOpen
  \bibfield  {author} {\bibinfo {author} {\bibfnamefont {A.}~\bibnamefont
  {Bianconi}}\ and\ \bibinfo {author} {\bibfnamefont {E.}~\bibnamefont
  {Tomasi-Gustafsson}},\ }\href {\doibase 10.1103/PhysRevLett.114.232301}
  {\bibfield  {journal} {\bibinfo  {journal} {Phys. Rev. Lett.}\ }\textbf
  {\bibinfo {volume} {114}},\ \bibinfo {pages} {232301} (\bibinfo {year}
  {2015})},\ \Eprint {http://arxiv.org/abs/1503.02140} {arXiv:1503.02140
  [nucl-th]} \BibitemShut {NoStop}%
\bibitem [{\citenamefont {Zhou}\ \emph {et~al.}(2009)\citenamefont {Zhou},
  \citenamefont {Chen},\ and\ \citenamefont {Dong}}]{Zhou:2009xb}%
  \BibitemOpen
  \bibfield  {author} {\bibinfo {author} {\bibfnamefont {H.~Q.}\ \bibnamefont
  {Zhou}}, \bibinfo {author} {\bibfnamefont {D.~Y.}\ \bibnamefont {Chen}}, \
  and\ \bibinfo {author} {\bibfnamefont {Y.~B.}\ \bibnamefont {Dong}},\ }\href
  {\doibase 10.1016/j.physletb.2009.04.029} {\bibfield  {journal} {\bibinfo
  {journal} {Phys. Lett. B}\ }\textbf {\bibinfo {volume} {675}},\ \bibinfo
  {pages} {305} (\bibinfo {year} {2009})},\ \Eprint
  {http://arxiv.org/abs/0903.0301} {arXiv:0903.0301 [hep-ph]} \BibitemShut
  {NoStop}%
\bibitem [{\citenamefont {Chen}\ \emph {et~al.}(2009)\citenamefont {Chen},
  \citenamefont {Zhou},\ and\ \citenamefont {Dong}}]{Chen:2009ze}%
  \BibitemOpen
  \bibfield  {author} {\bibinfo {author} {\bibfnamefont {D.-Y.}\ \bibnamefont
  {Chen}}, \bibinfo {author} {\bibfnamefont {H.-Q.}\ \bibnamefont {Zhou}}, \
  and\ \bibinfo {author} {\bibfnamefont {Y.-B.}\ \bibnamefont {Dong}},\ }\href
  {\doibase 10.1088/1674-1137/33/12/049} {\bibfield  {journal} {\bibinfo
  {journal} {Chin. Phys. C}\ }\textbf {\bibinfo {volume} {33}},\ \bibinfo
  {pages} {1336} (\bibinfo {year} {2009})}\BibitemShut {NoStop}%
\bibitem [{\citenamefont {Zhou}(2010)}]{Zhou:2010zzt}%
  \BibitemOpen
  \bibfield  {author} {\bibinfo {author} {\bibfnamefont {H.-Q.}\ \bibnamefont
  {Zhou}},\ }\href {\doibase 10.1088/1674-1137/34/6/042} {\bibfield  {journal}
  {\bibinfo  {journal} {Chin. Phys. C}\ }\textbf {\bibinfo {volume} {34}},\
  \bibinfo {pages} {869} (\bibinfo {year} {2010})}\BibitemShut {NoStop}%
\bibitem [{\citenamefont {Zhou}\ and\ \citenamefont {Zou}(2012)}]{Zhou:2011yz}%
  \BibitemOpen
  \bibfield  {author} {\bibinfo {author} {\bibfnamefont {H.-Q.}\ \bibnamefont
  {Zhou}}\ and\ \bibinfo {author} {\bibfnamefont {B.-S.}\ \bibnamefont {Zou}},\
  }\href {\doibase 10.1016/j.nuclphysa.2012.04.004} {\bibfield  {journal}
  {\bibinfo  {journal} {Nucl. Phys. A}\ }\textbf {\bibinfo {volume} {883}},\
  \bibinfo {pages} {49} (\bibinfo {year} {2012})},\ \Eprint
  {http://arxiv.org/abs/1112.4615} {arXiv:1112.4615 [hep-ph]} \BibitemShut
  {NoStop}%
\bibitem [{\citenamefont {Borisyuk}\ and\ \citenamefont
  {Kobushkin}(2008)}]{Borisyuk:2008es}%
  \BibitemOpen
  \bibfield  {author} {\bibinfo {author} {\bibfnamefont {D.}~\bibnamefont
  {Borisyuk}}\ and\ \bibinfo {author} {\bibfnamefont {A.}~\bibnamefont
  {Kobushkin}},\ }\href {\doibase 10.1103/PhysRevC.78.025208} {\bibfield
  {journal} {\bibinfo  {journal} {Phys. Rev. C}\ }\textbf {\bibinfo {volume}
  {78}},\ \bibinfo {pages} {025208} (\bibinfo {year} {2008})},\ \Eprint
  {http://arxiv.org/abs/0804.4128} {arXiv:0804.4128 [nucl-th]} \BibitemShut
  {NoStop}%
\bibitem [{\citenamefont {Lin}\ \emph {et~al.}(2022{\natexlab{a}})\citenamefont
  {Lin}, \citenamefont {Hammer},\ and\ \citenamefont
  {Mei{\ss}ner}}]{Lin:2021xrc}%
  \BibitemOpen
  \bibfield  {author} {\bibinfo {author} {\bibfnamefont {Y.-H.}\ \bibnamefont
  {Lin}}, \bibinfo {author} {\bibfnamefont {H.-W.}\ \bibnamefont {Hammer}}, \
  and\ \bibinfo {author} {\bibfnamefont {U.-G.}\ \bibnamefont {Mei{\ss}ner}},\
  }\href {\doibase 10.1103/PhysRevLett.128.052002} {\bibfield  {journal}
  {\bibinfo  {journal} {Phys. Rev. Lett.}\ }\textbf {\bibinfo {volume} {128}},\
  \bibinfo {pages} {052002} (\bibinfo {year} {2022}{\natexlab{a}})},\ \Eprint
  {http://arxiv.org/abs/2109.12961} {arXiv:2109.12961 [hep-ph]} \BibitemShut
  {NoStop}%
\bibitem [{\citenamefont {Lin}\ \emph {et~al.}(2022{\natexlab{b}})\citenamefont
  {Lin}, \citenamefont {Hammer},\ and\ \citenamefont
  {Mei{\ss}ner}}]{Lin:2022baj}%
  \BibitemOpen
  \bibfield  {author} {\bibinfo {author} {\bibfnamefont {Y.-H.}\ \bibnamefont
  {Lin}}, \bibinfo {author} {\bibfnamefont {H.-W.}\ \bibnamefont {Hammer}}, \
  and\ \bibinfo {author} {\bibfnamefont {U.-G.}\ \bibnamefont {Mei{\ss}ner}},\
  }\href {\doibase 10.1140/epjc/s10052-022-11056-8} {\bibfield  {journal}
  {\bibinfo  {journal} {Eur. Phys. J. C}\ }\textbf {\bibinfo {volume} {82}},\
  \bibinfo {pages} {1091} (\bibinfo {year} {2022}{\natexlab{b}})},\ \Eprint
  {http://arxiv.org/abs/2208.14802} {arXiv:2208.14802 [hep-ph]} \BibitemShut
  {NoStop}%
\bibitem [{\citenamefont {Lin}\ \emph {et~al.}(2023)\citenamefont {Lin},
  \citenamefont {Hammer},\ and\ \citenamefont {Mei{\ss}ner}}]{Lin:2022dyu}%
  \BibitemOpen
  \bibfield  {author} {\bibinfo {author} {\bibfnamefont {Y.-H.}\ \bibnamefont
  {Lin}}, \bibinfo {author} {\bibfnamefont {H.-W.}\ \bibnamefont {Hammer}}, \
  and\ \bibinfo {author} {\bibfnamefont {U.-G.}\ \bibnamefont {Mei{\ss}ner}},\
  }\href {\doibase 10.1140/epja/s10050-023-00973-1} {\bibfield  {journal}
  {\bibinfo  {journal} {Eur. Phys. J. A}\ }\textbf {\bibinfo {volume} {59}},\
  \bibinfo {pages} {54} (\bibinfo {year} {2023})},\ \Eprint
  {http://arxiv.org/abs/2205.00850} {arXiv:2205.00850 [hep-ph]} \BibitemShut
  {NoStop}%
\bibitem [{\citenamefont {Tomasi-Gustafsson}\ \emph {et~al.}(2008)\citenamefont
  {Tomasi-Gustafsson}, \citenamefont {Kuraev}, \citenamefont {Bakmaev},\ and\
  \citenamefont {Pacetti}}]{Tomasi-Gustafsson:2007vpn}%
  \BibitemOpen
  \bibfield  {author} {\bibinfo {author} {\bibfnamefont {E.}~\bibnamefont
  {Tomasi-Gustafsson}}, \bibinfo {author} {\bibfnamefont {E.~A.}\ \bibnamefont
  {Kuraev}}, \bibinfo {author} {\bibfnamefont {S.}~\bibnamefont {Bakmaev}}, \
  and\ \bibinfo {author} {\bibfnamefont {S.}~\bibnamefont {Pacetti}},\ }\href
  {\doibase 10.1016/j.physletb.2007.11.031} {\bibfield  {journal} {\bibinfo
  {journal} {Phys. Lett. B}\ }\textbf {\bibinfo {volume} {659}},\ \bibinfo
  {pages} {197} (\bibinfo {year} {2008})},\ \Eprint
  {http://arxiv.org/abs/0710.0454} {arXiv:0710.0454 [hep-ph]} \BibitemShut
  {NoStop}%
\bibitem [{\citenamefont {Tomasi-Gustafsson}\ and\ \citenamefont
  {Gakh}(2005)}]{Tomasi-Gustafsson:2004fss}%
  \BibitemOpen
  \bibfield  {author} {\bibinfo {author} {\bibfnamefont {E.}~\bibnamefont
  {Tomasi-Gustafsson}}\ and\ \bibinfo {author} {\bibfnamefont {G.~I.}\
  \bibnamefont {Gakh}},\ }\href {\doibase 10.1103/PhysRevC.72.015209}
  {\bibfield  {journal} {\bibinfo  {journal} {Phys. Rev. C}\ }\textbf {\bibinfo
  {volume} {72}},\ \bibinfo {pages} {015209} (\bibinfo {year} {2005})},\
  \Eprint {http://arxiv.org/abs/hep-ph/0412137} {arXiv:hep-ph/0412137}
  \BibitemShut {NoStop}%
\bibitem [{\citenamefont {Xia}(2026)}]{Xia:2025rio}%
  \BibitemOpen
  \bibfield  {author} {\bibinfo {author} {\bibfnamefont {L.}~\bibnamefont
  {Xia}} (\bibinfo {collaboration} {BESIII}),\ }\href {\doibase
  10.22323/1.500.0097} {\bibfield  {journal} {\bibinfo  {journal} {PoS}\
  }\textbf {\bibinfo {volume} {HADRON2025}},\ \bibinfo {pages} {097} (\bibinfo
  {year} {2026})},\ \Eprint {http://arxiv.org/abs/2508.12812} {arXiv:2508.12812
  [hep-ex]} \BibitemShut {NoStop}%
\bibitem [{\citenamefont {Gao}\ \emph {et~al.}(2026)\citenamefont {Gao},
  \citenamefont {Ping},\ and\ \citenamefont {Zhao}}]{Gao:2026axl}%
  \BibitemOpen
  \bibfield  {author} {\bibinfo {author} {\bibfnamefont {Z.}~\bibnamefont
  {Gao}}, \bibinfo {author} {\bibfnamefont {R.}~\bibnamefont {Ping}}, \ and\
  \bibinfo {author} {\bibfnamefont {M.}~\bibnamefont {Zhao}},\ }\href@noop {}
  {\  (\bibinfo {year} {2026})},\ \Eprint {http://arxiv.org/abs/2602.20526}
  {arXiv:2602.20526 [hep-ph]} \BibitemShut {NoStop}%
\bibitem [{\citenamefont {Ablikim}\ \emph
  {et~al.}(2019{\natexlab{a}})\citenamefont {Ablikim} \emph
  {et~al.}}]{BESIII:2019nep}%
  \BibitemOpen
  \bibfield  {author} {\bibinfo {author} {\bibfnamefont {M.}~\bibnamefont
  {Ablikim}} \emph {et~al.} (\bibinfo {collaboration} {BESIII}),\ }\href
  {\doibase 10.1103/PhysRevLett.123.122003} {\bibfield  {journal} {\bibinfo
  {journal} {Phys. Rev. Lett.}\ }\textbf {\bibinfo {volume} {123}},\ \bibinfo
  {pages} {122003} (\bibinfo {year} {2019}{\natexlab{a}})},\ \Eprint
  {http://arxiv.org/abs/1903.09421} {arXiv:1903.09421 [hep-ex]} \BibitemShut
  {NoStop}%
\bibitem [{\citenamefont {Korner}\ and\ \citenamefont
  {Kuroda}(1977)}]{Korner:1976hv}%
  \BibitemOpen
  \bibfield  {author} {\bibinfo {author} {\bibfnamefont {J.~G.}\ \bibnamefont
  {Korner}}\ and\ \bibinfo {author} {\bibfnamefont {M.}~\bibnamefont
  {Kuroda}},\ }\href {\doibase 10.1103/PhysRevD.16.2165} {\bibfield  {journal}
  {\bibinfo  {journal} {Phys. Rev. D}\ }\textbf {\bibinfo {volume} {16}},\
  \bibinfo {pages} {2165} (\bibinfo {year} {1977})}\BibitemShut {NoStop}%
\bibitem [{\citenamefont {Claudson}\ \emph {et~al.}(1982)\citenamefont
  {Claudson}, \citenamefont {Glashow},\ and\ \citenamefont
  {Wise}}]{Claudson:1981fj}%
  \BibitemOpen
  \bibfield  {author} {\bibinfo {author} {\bibfnamefont {M.}~\bibnamefont
  {Claudson}}, \bibinfo {author} {\bibfnamefont {S.~L.}\ \bibnamefont
  {Glashow}}, \ and\ \bibinfo {author} {\bibfnamefont {M.~B.}\ \bibnamefont
  {Wise}},\ }\href {\doibase 10.1103/PhysRevD.25.1345} {\bibfield  {journal}
  {\bibinfo  {journal} {Phys. Rev. D}\ }\textbf {\bibinfo {volume} {25}},\
  \bibinfo {pages} {1345} (\bibinfo {year} {1982})}\BibitemShut {NoStop}%
\bibitem [{\citenamefont {Ablikim}\ \emph {et~al.}(2012)\citenamefont {Ablikim}
  \emph {et~al.}}]{BESIII:2012xdg}%
  \BibitemOpen
  \bibfield  {author} {\bibinfo {author} {\bibfnamefont {M.}~\bibnamefont
  {Ablikim}} \emph {et~al.} (\bibinfo {collaboration} {BESIII}),\ }\href
  {\doibase 10.1103/PhysRevD.86.032008} {\bibfield  {journal} {\bibinfo
  {journal} {Phys. Rev. D}\ }\textbf {\bibinfo {volume} {86}},\ \bibinfo
  {pages} {032008} (\bibinfo {year} {2012})},\ \Eprint
  {http://arxiv.org/abs/1207.1201} {arXiv:1207.1201 [hep-ex]} \BibitemShut
  {NoStop}%
\bibitem [{\citenamefont {Ablikim}\ \emph
  {et~al.}(2024{\natexlab{a}})\citenamefont {Ablikim} \emph
  {et~al.}}]{BESIII:2023pfv}%
  \BibitemOpen
  \bibfield  {author} {\bibinfo {author} {\bibfnamefont {M.}~\bibnamefont
  {Ablikim}} \emph {et~al.} (\bibinfo {collaboration} {BESIII}),\ }\href
  {\doibase 10.1103/PhysRevD.109.012002} {\bibfield  {journal} {\bibinfo
  {journal} {Phys. Rev. D}\ }\textbf {\bibinfo {volume} {109}},\ \bibinfo
  {pages} {012002} (\bibinfo {year} {2024}{\natexlab{a}})},\ \Eprint
  {http://arxiv.org/abs/2308.03361} {arXiv:2308.03361 [hep-ex]} \BibitemShut
  {NoStop}%
\bibitem [{\citenamefont {Ablikim}\ \emph
  {et~al.}(2024{\natexlab{b}})\citenamefont {Ablikim} \emph
  {et~al.}}]{BESIII:2023cvk}%
  \BibitemOpen
  \bibfield  {author} {\bibinfo {author} {\bibfnamefont {M.}~\bibnamefont
  {Ablikim}} \emph {et~al.} (\bibinfo {collaboration} {BESIII}),\ }\href
  {\doibase 10.1038/s41467-024-51802-y} {\bibfield  {journal} {\bibinfo
  {journal} {Nature Commun.}\ }\textbf {\bibinfo {volume} {15}},\ \bibinfo
  {pages} {8812} (\bibinfo {year} {2024}{\natexlab{b}})},\ \Eprint
  {http://arxiv.org/abs/2309.04139} {arXiv:2309.04139 [hep-ex]} \BibitemShut
  {NoStop}%
\bibitem [{\citenamefont {Zhu}\ \emph {et~al.}(2015)\citenamefont {Zhu},
  \citenamefont {Mo},\ and\ \citenamefont {Yuan}}]{Zhu:2015bha}%
  \BibitemOpen
  \bibfield  {author} {\bibinfo {author} {\bibfnamefont {K.}~\bibnamefont
  {Zhu}}, \bibinfo {author} {\bibfnamefont {X.-H.}\ \bibnamefont {Mo}}, \ and\
  \bibinfo {author} {\bibfnamefont {C.-Z.}\ \bibnamefont {Yuan}},\ }\href
  {\doibase 10.1142/S0217751X15501481} {\bibfield  {journal} {\bibinfo
  {journal} {Int. J. Mod. Phys. A}\ }\textbf {\bibinfo {volume} {30}},\
  \bibinfo {pages} {1550148} (\bibinfo {year} {2015})},\ \Eprint
  {http://arxiv.org/abs/1505.03930} {arXiv:1505.03930 [hep-ph]} \BibitemShut
  {NoStop}%
\bibitem [{\citenamefont {Baldini~Ferroli}\ \emph {et~al.}(2019)\citenamefont
  {Baldini~Ferroli}, \citenamefont {Mangoni}, \citenamefont {Pacetti},\ and\
  \citenamefont {Zhu}}]{BaldiniFerroli:2019abd}%
  \BibitemOpen
  \bibfield  {author} {\bibinfo {author} {\bibfnamefont {R.}~\bibnamefont
  {Baldini~Ferroli}}, \bibinfo {author} {\bibfnamefont {A.}~\bibnamefont
  {Mangoni}}, \bibinfo {author} {\bibfnamefont {S.}~\bibnamefont {Pacetti}}, \
  and\ \bibinfo {author} {\bibfnamefont {K.}~\bibnamefont {Zhu}},\ }\href
  {\doibase 10.1016/j.physletb.2019.135041} {\bibfield  {journal} {\bibinfo
  {journal} {Phys. Lett. B}\ }\textbf {\bibinfo {volume} {799}},\ \bibinfo
  {pages} {135041} (\bibinfo {year} {2019})},\ \Eprint
  {http://arxiv.org/abs/1905.01069} {arXiv:1905.01069 [hep-ph]} \BibitemShut
  {NoStop}%
\bibitem [{\citenamefont {Ferroli}\ \emph {et~al.}(2020)\citenamefont
  {Ferroli}, \citenamefont {Mangoni},\ and\ \citenamefont
  {Pacetti}}]{Ferroli:2020xnv}%
  \BibitemOpen
  \bibfield  {author} {\bibinfo {author} {\bibfnamefont {R.~B.}\ \bibnamefont
  {Ferroli}}, \bibinfo {author} {\bibfnamefont {A.}~\bibnamefont {Mangoni}}, \
  and\ \bibinfo {author} {\bibfnamefont {S.}~\bibnamefont {Pacetti}},\ }\href
  {\doibase 10.1140/epjc/s10052-020-08474-x} {\bibfield  {journal} {\bibinfo
  {journal} {Eur. Phys. J. C}\ }\textbf {\bibinfo {volume} {80}},\ \bibinfo
  {pages} {903} (\bibinfo {year} {2020})},\ \Eprint
  {http://arxiv.org/abs/2007.12380} {arXiv:2007.12380 [hep-ph]} \BibitemShut
  {NoStop}%
\bibitem [{\citenamefont {Mo}\ and\ \citenamefont {Zhang}(2022)}]{Mo:2021asa}%
  \BibitemOpen
  \bibfield  {author} {\bibinfo {author} {\bibfnamefont {X.~H.}\ \bibnamefont
  {Mo}}\ and\ \bibinfo {author} {\bibfnamefont {J.~Y.}\ \bibnamefont {Zhang}},\
  }\href {\doibase 10.1016/j.physletb.2022.136927} {\bibfield  {journal}
  {\bibinfo  {journal} {Phys. Lett. B}\ }\textbf {\bibinfo {volume} {826}},\
  \bibinfo {pages} {136927} (\bibinfo {year} {2022})},\ \Eprint
  {http://arxiv.org/abs/2111.15045} {arXiv:2111.15045 [hep-ph]} \BibitemShut
  {NoStop}%
\bibitem [{\citenamefont {Mo}\ \emph {et~al.}(2023)\citenamefont {Mo},
  \citenamefont {Wang},\ and\ \citenamefont {Zhang}}]{Mo:2023wrf}%
  \BibitemOpen
  \bibfield  {author} {\bibinfo {author} {\bibfnamefont {X.~H.}\ \bibnamefont
  {Mo}}, \bibinfo {author} {\bibfnamefont {P.}~\bibnamefont {Wang}}, \ and\
  \bibinfo {author} {\bibfnamefont {J.~Y.}\ \bibnamefont {Zhang}},\ }\href
  {\doibase 10.1103/PhysRevD.107.094009} {\bibfield  {journal} {\bibinfo
  {journal} {Phys. Rev. D}\ }\textbf {\bibinfo {volume} {107}},\ \bibinfo
  {pages} {094009} (\bibinfo {year} {2023})},\ \Eprint
  {http://arxiv.org/abs/2303.12235} {arXiv:2303.12235 [hep-ph]} \BibitemShut
  {NoStop}%
\bibitem [{\citenamefont {Geng}\ \emph {et~al.}(2023)\citenamefont {Geng},
  \citenamefont {Liu},\ and\ \citenamefont {Zhang}}]{Geng:2023yqo}%
  \BibitemOpen
  \bibfield  {author} {\bibinfo {author} {\bibfnamefont {C.-Q.}\ \bibnamefont
  {Geng}}, \bibinfo {author} {\bibfnamefont {C.-W.}\ \bibnamefont {Liu}}, \
  and\ \bibinfo {author} {\bibfnamefont {J.}~\bibnamefont {Zhang}},\ }\href
  {\doibase 10.1140/epjc/s10052-023-12099-1} {\bibfield  {journal} {\bibinfo
  {journal} {Eur. Phys. J. C}\ }\textbf {\bibinfo {volume} {83}},\ \bibinfo
  {pages} {973} (\bibinfo {year} {2023})},\ \Eprint
  {http://arxiv.org/abs/2306.02138} {arXiv:2306.02138 [hep-ph]} \BibitemShut
  {NoStop}%
\bibitem [{\citenamefont {Rosini}\ and\ \citenamefont
  {Pacetti}(2025)}]{Rosini:2025wfm}%
  \BibitemOpen
  \bibfield  {author} {\bibinfo {author} {\bibfnamefont {F.}~\bibnamefont
  {Rosini}}\ and\ \bibinfo {author} {\bibfnamefont {S.}~\bibnamefont
  {Pacetti}},\ }\href {\doibase 10.1140/epjc/s10052-025-13976-7} {\bibfield
  {journal} {\bibinfo  {journal} {Eur. Phys. J. C}\ }\textbf {\bibinfo {volume}
  {85}},\ \bibinfo {pages} {236} (\bibinfo {year} {2025})},\ \Eprint
  {http://arxiv.org/abs/2501.04449} {arXiv:2501.04449 [hep-ph]} \BibitemShut
  {NoStop}%
\bibitem [{\citenamefont {Chen}\ \emph
  {et~al.}(2025{\natexlab{b}})\citenamefont {Chen}, \citenamefont {Yan},
  \citenamefont {Long},\ and\ \citenamefont {Xie}}]{Chen:2025xci}%
  \BibitemOpen
  \bibfield  {author} {\bibinfo {author} {\bibfnamefont {C.}~\bibnamefont
  {Chen}}, \bibinfo {author} {\bibfnamefont {B.}~\bibnamefont {Yan}}, \bibinfo
  {author} {\bibfnamefont {B.-W.}\ \bibnamefont {Long}}, \ and\ \bibinfo
  {author} {\bibfnamefont {J.-J.}\ \bibnamefont {Xie}},\ }\href {\doibase
  10.1088/0256-307X/42/12/120201} {\bibfield  {journal} {\bibinfo  {journal}
  {Chin. Phys. Lett.}\ }\textbf {\bibinfo {volume} {42}},\ \bibinfo {pages}
  {120201} (\bibinfo {year} {2025}{\natexlab{b}})},\ \Eprint
  {http://arxiv.org/abs/2507.01277} {arXiv:2507.01277 [hep-ph]} \BibitemShut
  {NoStop}%
\bibitem [{\citenamefont {Zhang}\ \emph {et~al.}(2025)\citenamefont {Zhang},
  \citenamefont {Liu}, \citenamefont {Ping}, \citenamefont {Song},\ and\
  \citenamefont {Yang}}]{Zhang:2025oks}%
  \BibitemOpen
  \bibfield  {author} {\bibinfo {author} {\bibfnamefont {Z.}~\bibnamefont
  {Zhang}}, \bibinfo {author} {\bibfnamefont {T.}~\bibnamefont {Liu}}, \bibinfo
  {author} {\bibfnamefont {R.-G.}\ \bibnamefont {Ping}}, \bibinfo {author}
  {\bibfnamefont {J.~J.}\ \bibnamefont {Song}}, \ and\ \bibinfo {author}
  {\bibfnamefont {W.}~\bibnamefont {Yang}},\ }\href {\doibase
  10.1103/mzkm-qw8w} {\bibfield  {journal} {\bibinfo  {journal} {Phys. Rev. D}\
  }\textbf {\bibinfo {volume} {112}},\ \bibinfo {pages} {096012} (\bibinfo
  {year} {2025})},\ \Eprint {http://arxiv.org/abs/2508.01813} {arXiv:2508.01813
  [hep-ph]} \BibitemShut {NoStop}%
\bibitem [{\citenamefont {Galster}\ \emph {et~al.}(1971)\citenamefont
  {Galster}, \citenamefont {Klein}, \citenamefont {Moritz}, \citenamefont
  {Schmidt}, \citenamefont {Wegener},\ and\ \citenamefont
  {Bleckwenn}}]{Galster:1971kv}%
  \BibitemOpen
  \bibfield  {author} {\bibinfo {author} {\bibfnamefont {S.}~\bibnamefont
  {Galster}}, \bibinfo {author} {\bibfnamefont {H.}~\bibnamefont {Klein}},
  \bibinfo {author} {\bibfnamefont {J.}~\bibnamefont {Moritz}}, \bibinfo
  {author} {\bibfnamefont {K.~H.}\ \bibnamefont {Schmidt}}, \bibinfo {author}
  {\bibfnamefont {D.}~\bibnamefont {Wegener}}, \ and\ \bibinfo {author}
  {\bibfnamefont {J.}~\bibnamefont {Bleckwenn}},\ }\href {\doibase
  10.1016/0550-3213(71)90068-X} {\bibfield  {journal} {\bibinfo  {journal}
  {Nucl. Phys. B}\ }\textbf {\bibinfo {volume} {32}},\ \bibinfo {pages} {221}
  (\bibinfo {year} {1971})}\BibitemShut {NoStop}%
\bibitem [{\citenamefont {Kelly}(2004)}]{Kelly:2004hm}%
  \BibitemOpen
  \bibfield  {author} {\bibinfo {author} {\bibfnamefont {J.~J.}\ \bibnamefont
  {Kelly}},\ }\href {\doibase 10.1103/PhysRevC.70.068202} {\bibfield  {journal}
  {\bibinfo  {journal} {Phys. Rev. C}\ }\textbf {\bibinfo {volume} {70}},\
  \bibinfo {pages} {068202} (\bibinfo {year} {2004})}\BibitemShut {NoStop}%
\bibitem [{\citenamefont {Alberico}\ \emph {et~al.}(2009)\citenamefont
  {Alberico}, \citenamefont {Bilenky}, \citenamefont {Giunti},\ and\
  \citenamefont {Graczyk}}]{Alberico:2008sz}%
  \BibitemOpen
  \bibfield  {author} {\bibinfo {author} {\bibfnamefont {W.~M.}\ \bibnamefont
  {Alberico}}, \bibinfo {author} {\bibfnamefont {S.~M.}\ \bibnamefont
  {Bilenky}}, \bibinfo {author} {\bibfnamefont {C.}~\bibnamefont {Giunti}}, \
  and\ \bibinfo {author} {\bibfnamefont {K.~M.}\ \bibnamefont {Graczyk}},\
  }\href {\doibase 10.1103/PhysRevC.79.065204} {\bibfield  {journal} {\bibinfo
  {journal} {Phys. Rev. C}\ }\textbf {\bibinfo {volume} {79}},\ \bibinfo
  {pages} {065204} (\bibinfo {year} {2009})},\ \Eprint
  {http://arxiv.org/abs/0812.3539} {arXiv:0812.3539 [hep-ph]} \BibitemShut
  {NoStop}%
\bibitem [{\citenamefont {Navas}\ \emph {et~al.}(2024)\citenamefont {Navas}
  \emph {et~al.}}]{ParticleDataGroup:2024cfk}%
  \BibitemOpen
  \bibfield  {author} {\bibinfo {author} {\bibfnamefont {S.}~\bibnamefont
  {Navas}} \emph {et~al.} (\bibinfo {collaboration} {Particle Data Group}),\
  }\href {\doibase 10.1103/PhysRevD.110.030001} {\bibfield  {journal} {\bibinfo
   {journal} {Phys. Rev. D}\ }\textbf {\bibinfo {volume} {110}},\ \bibinfo
  {pages} {030001} (\bibinfo {year} {2024})}\BibitemShut {NoStop}%
\bibitem [{\citenamefont {Ablikim}\ \emph {et~al.}(2026)\citenamefont {Ablikim}
  \emph {et~al.}}]{BESIII:2026ubj}%
  \BibitemOpen
  \bibfield  {author} {\bibinfo {author} {\bibfnamefont {M.}~\bibnamefont
  {Ablikim}} \emph {et~al.} (\bibinfo {collaboration} {BESIII}),\ }\href@noop
  {} {\  (\bibinfo {year} {2026})},\ \Eprint {http://arxiv.org/abs/2608.16076}
  {arXiv:2608.16076 [hep-ex]} \BibitemShut {NoStop}%
\bibitem [{\citenamefont {Geng}\ \emph {et~al.}(2008)\citenamefont {Geng},
  \citenamefont {Martin~Camalich}, \citenamefont {Alvarez-Ruso},\ and\
  \citenamefont {Vicente~Vacas}}]{Geng:2008mf}%
  \BibitemOpen
  \bibfield  {author} {\bibinfo {author} {\bibfnamefont {L.~S.}\ \bibnamefont
  {Geng}}, \bibinfo {author} {\bibfnamefont {J.}~\bibnamefont
  {Martin~Camalich}}, \bibinfo {author} {\bibfnamefont {L.}~\bibnamefont
  {Alvarez-Ruso}}, \ and\ \bibinfo {author} {\bibfnamefont {M.~J.}\
  \bibnamefont {Vicente~Vacas}},\ }\href {\doibase
  10.1103/PhysRevLett.101.222002} {\bibfield  {journal} {\bibinfo  {journal}
  {Phys. Rev. Lett.}\ }\textbf {\bibinfo {volume} {101}},\ \bibinfo {pages}
  {222002} (\bibinfo {year} {2008})},\ \Eprint {http://arxiv.org/abs/0805.1419}
  {arXiv:0805.1419 [hep-ph]} \BibitemShut {NoStop}%
\bibitem [{\citenamefont {Flores-Mendieta}\ \emph {et~al.}(2021)\citenamefont
  {Flores-Mendieta}, \citenamefont {Garcia}, \citenamefont {Hernandez},\ and\
  \citenamefont {Trejo}}]{Flores-Mendieta:2021yzz}%
  \BibitemOpen
  \bibfield  {author} {\bibinfo {author} {\bibfnamefont {R.}~\bibnamefont
  {Flores-Mendieta}}, \bibinfo {author} {\bibfnamefont {C.~I.}\ \bibnamefont
  {Garcia}}, \bibinfo {author} {\bibfnamefont {J.}~\bibnamefont {Hernandez}}, \
  and\ \bibinfo {author} {\bibfnamefont {M.~A.}\ \bibnamefont {Trejo}},\ }\href
  {\doibase 10.1103/PhysRevD.104.114024} {\bibfield  {journal} {\bibinfo
  {journal} {Phys. Rev. D}\ }\textbf {\bibinfo {volume} {104}},\ \bibinfo
  {pages} {114024} (\bibinfo {year} {2021})},\ \Eprint
  {http://arxiv.org/abs/2109.04195} {arXiv:2109.04195 [hep-ph]} \BibitemShut
  {NoStop}%
\bibitem [{\citenamefont {Mertig}\ \emph {et~al.}(1991)\citenamefont {Mertig},
  \citenamefont {Bohm},\ and\ \citenamefont {Denner}}]{Mertig:1990an}%
  \BibitemOpen
  \bibfield  {author} {\bibinfo {author} {\bibfnamefont {R.}~\bibnamefont
  {Mertig}}, \bibinfo {author} {\bibfnamefont {M.}~\bibnamefont {Bohm}}, \ and\
  \bibinfo {author} {\bibfnamefont {A.}~\bibnamefont {Denner}},\ }\href
  {\doibase 10.1016/0010-4655(91)90130-D} {\bibfield  {journal} {\bibinfo
  {journal} {Comput. Phys. Commun.}\ }\textbf {\bibinfo {volume} {64}},\
  \bibinfo {pages} {345} (\bibinfo {year} {1991})}\BibitemShut {NoStop}%
\bibitem [{\citenamefont {Passarino}\ and\ \citenamefont
  {Veltman}(1979)}]{Passarino:1978jh}%
  \BibitemOpen
  \bibfield  {author} {\bibinfo {author} {\bibfnamefont {G.}~\bibnamefont
  {Passarino}}\ and\ \bibinfo {author} {\bibfnamefont {M.~J.~G.}\ \bibnamefont
  {Veltman}},\ }\href {\doibase 10.1016/0550-3213(79)90234-7} {\bibfield
  {journal} {\bibinfo  {journal} {Nucl. Phys. B}\ }\textbf {\bibinfo {volume}
  {160}},\ \bibinfo {pages} {151} (\bibinfo {year} {1979})}\BibitemShut
  {NoStop}%
\bibitem [{\citenamefont {Hahn}\ and\ \citenamefont
  {Perez-Victoria}(1999)}]{Hahn:1998yk}%
  \BibitemOpen
  \bibfield  {author} {\bibinfo {author} {\bibfnamefont {T.}~\bibnamefont
  {Hahn}}\ and\ \bibinfo {author} {\bibfnamefont {M.}~\bibnamefont
  {Perez-Victoria}},\ }\href {\doibase 10.1016/S0010-4655(98)00173-8}
  {\bibfield  {journal} {\bibinfo  {journal} {Comput. Phys. Commun.}\ }\textbf
  {\bibinfo {volume} {118}},\ \bibinfo {pages} {153} (\bibinfo {year}
  {1999})},\ \Eprint {http://arxiv.org/abs/hep-ph/9807565}
  {arXiv:hep-ph/9807565} \BibitemShut {NoStop}%
\bibitem [{\citenamefont {Ablikim}\ \emph {et~al.}(2021)\citenamefont {Ablikim}
  \emph {et~al.}}]{BESIII:2021mus}%
  \BibitemOpen
  \bibfield  {author} {\bibinfo {author} {\bibfnamefont {M.}~\bibnamefont
  {Ablikim}} \emph {et~al.} (\bibinfo {collaboration} {BESIII}),\ }\href
  {\doibase 10.1103/PhysRevD.103.112004} {\bibfield  {journal} {\bibinfo
  {journal} {Phys. Rev. D}\ }\textbf {\bibinfo {volume} {103}},\ \bibinfo
  {pages} {112004} (\bibinfo {year} {2021})},\ \Eprint
  {http://arxiv.org/abs/2103.16779} {arXiv:2103.16779 [hep-ex]} \BibitemShut
  {NoStop}%
\bibitem [{\citenamefont {Achasov}\ \emph {et~al.}(2024)\citenamefont {Achasov}
  \emph {et~al.}}]{Achasov:2023gey}%
  \BibitemOpen
  \bibfield  {author} {\bibinfo {author} {\bibfnamefont {M.}~\bibnamefont
  {Achasov}} \emph {et~al.},\ }\href {\doibase 10.1007/s11467-023-1333-z}
  {\bibfield  {journal} {\bibinfo  {journal} {Front. Phys. (Beijing)}\ }\textbf
  {\bibinfo {volume} {19}},\ \bibinfo {pages} {14701} (\bibinfo {year}
  {2024})},\ \Eprint {http://arxiv.org/abs/2303.15790} {arXiv:2303.15790
  [hep-ex]} \BibitemShut {NoStop}%
\bibitem [{\citenamefont {Ablikim}\ \emph
  {et~al.}(2019{\natexlab{b}})\citenamefont {Ablikim} \emph
  {et~al.}}]{BESIII:2018wid}%
  \BibitemOpen
  \bibfield  {author} {\bibinfo {author} {\bibfnamefont {M.}~\bibnamefont
  {Ablikim}} \emph {et~al.} (\bibinfo {collaboration} {BESIII}),\ }\href
  {\doibase 10.1016/j.physletb.2019.03.001} {\bibfield  {journal} {\bibinfo
  {journal} {Phys. Lett. B}\ }\textbf {\bibinfo {volume} {791}},\ \bibinfo
  {pages} {375} (\bibinfo {year} {2019}{\natexlab{b}})},\ \Eprint
  {http://arxiv.org/abs/1808.02166} {arXiv:1808.02166 [hep-ex]} \BibitemShut
  {NoStop}%
\bibitem [{\citenamefont {Ablikim}\ \emph {et~al.}(2025)\citenamefont {Ablikim}
  \emph {et~al.}}]{BESIII:2025cpv}%
  \BibitemOpen
  \bibfield  {author} {\bibinfo {author} {\bibfnamefont {M.}~\bibnamefont
  {Ablikim}} \emph {et~al.} (\bibinfo {collaboration} {BESIII}),\ }\href@noop
  {} {\  (\bibinfo {year} {2025})},\ \Eprint {http://arxiv.org/abs/2502.19850}
  {arXiv:2502.19850 [hep-ex]} \BibitemShut {NoStop}%
\bibitem [{\citenamefont {Carlson}\ and\ \citenamefont
  {Vanderhaeghen}(2007)}]{Carlson:2007sp}%
  \BibitemOpen
  \bibfield  {author} {\bibinfo {author} {\bibfnamefont {C.~E.}\ \bibnamefont
  {Carlson}}\ and\ \bibinfo {author} {\bibfnamefont {M.}~\bibnamefont
  {Vanderhaeghen}},\ }\href {\doibase 10.1146/annurev.nucl.57.090506.123116}
  {\bibfield  {journal} {\bibinfo  {journal} {Ann. Rev. Nucl. Part. Sci.}\
  }\textbf {\bibinfo {volume} {57}},\ \bibinfo {pages} {171} (\bibinfo {year}
  {2007})},\ \Eprint {http://arxiv.org/abs/hep-ph/0701272}
  {arXiv:hep-ph/0701272} \BibitemShut {NoStop}%
\bibitem [{\citenamefont {Aubert}\ \emph {et~al.}(2007)\citenamefont {Aubert}
  \emph {et~al.}}]{BaBar:2007fsu}%
  \BibitemOpen
  \bibfield  {author} {\bibinfo {author} {\bibfnamefont {B.}~\bibnamefont
  {Aubert}} \emph {et~al.} (\bibinfo {collaboration} {BaBar}),\ }\href
  {\doibase 10.1103/PhysRevD.76.092006} {\bibfield  {journal} {\bibinfo
  {journal} {Phys. Rev. D}\ }\textbf {\bibinfo {volume} {76}},\ \bibinfo
  {pages} {092006} (\bibinfo {year} {2007})},\ \Eprint
  {http://arxiv.org/abs/0709.1988} {arXiv:0709.1988 [hep-ex]} \BibitemShut
  {NoStop}%
\bibitem [{\citenamefont {Granados}\ \emph {et~al.}(2017)\citenamefont
  {Granados}, \citenamefont {Leupold},\ and\ \citenamefont
  {Perotti}}]{Granados:2017cib}%
  \BibitemOpen
  \bibfield  {author} {\bibinfo {author} {\bibfnamefont {C.}~\bibnamefont
  {Granados}}, \bibinfo {author} {\bibfnamefont {S.}~\bibnamefont {Leupold}}, \
  and\ \bibinfo {author} {\bibfnamefont {E.}~\bibnamefont {Perotti}},\ }\href
  {\doibase 10.1140/epja/i2017-12324-4} {\bibfield  {journal} {\bibinfo
  {journal} {Eur. Phys. J. A}\ }\textbf {\bibinfo {volume} {53}},\ \bibinfo
  {pages} {117} (\bibinfo {year} {2017})},\ \Eprint
  {http://arxiv.org/abs/1701.09130} {arXiv:1701.09130 [hep-ph]} \BibitemShut
  {NoStop}%
\bibitem [{\citenamefont {Gari}\ and\ \citenamefont
  {Krumpelmann}(1985)}]{Gari:1984ia}%
  \BibitemOpen
  \bibfield  {author} {\bibinfo {author} {\bibfnamefont {M.}~\bibnamefont
  {Gari}}\ and\ \bibinfo {author} {\bibfnamefont {W.}~\bibnamefont
  {Krumpelmann}},\ }\href {\doibase 10.1007/BF01415153} {\bibfield  {journal}
  {\bibinfo  {journal} {Z. Phys. A}\ }\textbf {\bibinfo {volume} {322}},\
  \bibinfo {pages} {689} (\bibinfo {year} {1985})}\BibitemShut {NoStop}%
\bibitem [{\citenamefont {Gari}\ and\ \citenamefont
  {Kaulfuss}(1984)}]{Gari:1984pq}%
  \BibitemOpen
  \bibfield  {author} {\bibinfo {author} {\bibfnamefont {M.}~\bibnamefont
  {Gari}}\ and\ \bibinfo {author} {\bibfnamefont {U.}~\bibnamefont
  {Kaulfuss}},\ }\href {\doibase 10.1016/0370-2693(84)91166-3} {\bibfield
  {journal} {\bibinfo  {journal} {Phys. Lett. B}\ }\textbf {\bibinfo {volume}
  {136}},\ \bibinfo {pages} {139} (\bibinfo {year} {1984})}\BibitemShut
  {NoStop}%
\bibitem [{\citenamefont {Lepage}\ and\ \citenamefont
  {Brodsky}(1980)}]{Lepage:1980fj}%
  \BibitemOpen
  \bibfield  {author} {\bibinfo {author} {\bibfnamefont {G.~P.}\ \bibnamefont
  {Lepage}}\ and\ \bibinfo {author} {\bibfnamefont {S.~J.}\ \bibnamefont
  {Brodsky}},\ }\href {\doibase 10.1103/PhysRevD.22.2157} {\bibfield  {journal}
  {\bibinfo  {journal} {Phys. Rev. D}\ }\textbf {\bibinfo {volume} {22}},\
  \bibinfo {pages} {2157} (\bibinfo {year} {1980})}\BibitemShut {NoStop}%
\end{thebibliography}%

\end{document}